\documentclass{article}
\usepackage{ieeetrantools}
\usepackage{subcaption}
\usepackage{graphicx}
\usepackage{spconf,amsmath,graphicx}
\usepackage{setspace}
\usepackage{tabularx,booktabs}
\usepackage{dingbat}
\usepackage{diagbox}
\usepackage{float}
\usepackage{multirow} 
\usepackage[hyphens]{url}
\usepackage[pagebackref,breaklinks,colorlinks]{hyperref}
\usepackage{xcolor}
\usepackage{colortbl}
\usepackage{amsfonts, amsmath, amsthm, amssymb}
\usepackage{subcaption}
\hypersetup{breaklinks=true}
\usepackage{tikz}
\usepackage{textcomp}
\usepackage{lipsum}
\usepackage{svg}
\usepackage{enumitem}
\usepackage[nameinlink]{cleveref}
\usepackage{amssymb}
\usepackage{pifont}
\usepackage{scalerel}
\usepackage{stackengine}
\usepackage{listings}
\usepackage{microtype}
\setlist[itemize]{noitemsep}

\newcommand\blfootnote[1]{%
  \begingroup
  \renewcommand\thefootnote{}\footnote{#1}%
  \addtocounter{footnote}{-1}%
  \endgroup
}
\lstdefinelanguage{PowerShell}{
	morekeywords={
		Add-Content,Add-PSSnapin,Clear-Content,Clear-History,Clear-Host,Clear-Item,Clear-ItemProperty,Clear-Variable,Compare-Object,Connect-PSSession,ConvertFrom-String,Convert-Path,Copy-Item,Copy-ItemProperty,Disable-PSBreakpoint,Disconnect-PSSession,Enable-PSBreakpoint,Enter-PSSession,Exit-PSSession,Export-Alias,Export-Csv,Export-PSSession,ForEach-Object,Format-Custom,Format-Hex,Format-List,Format-Table,Format-Wide,Get-Alias,Get-ChildItem,Get-Clipboard,Get-Command,Get-ComputerInfo,Get-Content,Get-History,Get-Item,Get-ItemProperty,Get-ItemPropertyValue,Get-Job,Get-Location,Get-Member,Get-Module,Get-Process,Get-PSBreakpoint,Get-PSCallStack,Get-PSDrive,Get-PSSession,Get-PSSnapin,Get-Service,Get-TimeZone,Get-Unique,Get-Variable,Get-WmiObject,Group-Object,help,Import-Alias,Import-Csv,Import-Module,Import-PSSession,Invoke-Command,Invoke-Expression,Invoke-History,Invoke-Item,Invoke-RestMethod,Invoke-WebRequest,Invoke-WmiMethod,Measure-Object,mkdir,Move-Item,Move-ItemProperty,New-Alias,New-Item,New-Module,New-PSDrive,New-PSSession,New-PSSessionConfigurationFile,New-Variable,Out-GridView,Out-Host,Out-Printer,Pop-Location,powershell_ise.exe,Push-Location,Receive-Job,Receive-PSSession,Remove-Item,Remove-ItemProperty,Remove-Job,Remove-Module,Remove-PSBreakpoint,Remove-PSDrive,Remove-PSSession,Remove-PSSnapin,Remove-Variable,Remove-WmiObject,Rename-Item,Rename-ItemProperty,Resolve-Path,Resume-Job,Select-Object,Select-String,Set-Alias,Set-Clipboard,Set-Content,Set-Item,Set-ItemProperty,Set-Location,Set-PSBreakpoint,Set-TimeZone,Set-Variable,Set-WmiInstance,Show-Command,Sort-Object,Start-Job,Start-Process,Start-Service,Start-Sleep,Stop-Job,Stop-Process,Stop-Service,Suspend-Job,Tee-Object,Trace-Command,Wait-Job,Where-Object,Write-Output
	},
	morekeywords={
		Add-AppxPackage,Add-AppxProvisionedPackage,Add-AppxVolume,Add-BitsFile,Add-CertificateEnrollmentPolicyServer,Add-Computer,Add-Content,Add-History,Add-JobTrigger,Add-KdsRootKey,Add-LocalGroupMember,Add-Member,Add-PSSnapin,Add-Type,Add-WindowsCapability,Add-WindowsDriver,Add-WindowsImage,Add-WindowsPackage,Checkpoint-Computer,Clear-Content,Clear-EventLog,Clear-History,Clear-Item,Clear-ItemProperty,Clear-KdsCache,Clear-RecycleBin,Clear-Tpm,Clear-Variable,Clear-WindowsCorruptMountPoint,Compare-Object,Complete-BitsTransfer,Complete-DtiagnosticTransaction,Complete-Transaction,Confirm-SecureBootUEFI,Connect-PSSession,Connect-WSMan,ConvertFrom-Csv,ConvertFrom-Json,ConvertFrom-SecureString,ConvertFrom-String,ConvertFrom-StringData,Convert-Path,Convert-String,ConvertTo-Csv,ConvertTo-Html,ConvertTo-Json,ConvertTo-ProcessMitigationPolicy,ConvertTo-SecureString,ConvertTo-TpmOwnerAuth,ConvertTo-Xml,Copy-Item,Copy-ItemProperty,Debug-Job,Debug-Process,Debug-Runspace,Disable-AppBackgroundTaskDiagnosticLog,Disable-ComputerRestore,Disable-JobTrigger,Disable-LocalUser,Disable-PSBreakpoint,Disable-PSRemoting,Disable-PSSessionConfiguration,Disable-RunspaceDebug,Disable-ScheduledJob,Disable-TlsCipherSuite,Disable-TlsEccCurve,Disable-TlsSessionTicketKey,Disable-TpmAutoProvisioning,Disable-WindowsErrorReporting,Disable-WindowsOptionalFeature,Disable-WSManCredSSP,Disconnect-PSSession,Disconnect-WSMan,Dismount-AppxVolume,Dismount-WindowsImage,Enable-AppBackgroundTaskDiagnosticLog,Enable-ComputerRestore,Enable-JobTrigger,Enable-LocalUser,Enable-PSBreakpoint,Enable-PSRemoting,Enable-PSSessionConfiguration,Enable-RunspaceDebug,Enable-ScheduledJob,Enable-TlsCipherSuite,Enable-TlsEccCurve,Enable-TlsSessionTicketKey,Enable-TpmAutoProvisioning,Enable-WindowsErrorReporting,Enable-WindowsOptionalFeature,Enable-WSManCredSSP,Enter-PSHostProcess,Enter-PSSession,Exit-PSHostProcess,Exit-PSSession,Expand-WindowsCustomDataImage,Expand-WindowsImage,Export-Alias,Export-BinaryMiLog,Export-Certificate,Export-Clixml,Export-Console,Export-Counter,Export-Csv,Export-FormatData,Export-ModuleMember,Export-PfxCertificate,Export-ProvisioningPackage,Export-PSSession,Export-StartLayout,Export-StartLayoutEdgeAssets,Export-TlsSessionTicketKey,Export-Trace,Export-WindowsCapabilitySource,Export-WindowsDriver,Export-WindowsImage,Find-Package,Find-PackageProvider,ForEach-Object,Format-Custom,Format-List,Format-SecureBootUEFI,Format-Table,Format-Wide,Get-Acl,Get-Alias,Get-AppxDefaultVolume,Get-AppxPackage,Get-AppxPackageManifest,Get-AppxProvisionedPackage,Get-AppxVolume,Get-AuthenticodeSignature,Get-BitsTransfer,Get-Certificate,Get-CertificateAutoEnrollmentPolicy,Get-CertificateEnrollmentPolicyServer,Get-CertificateNotificationTask,Get-ChildItem,Get-CimAssociatedInstance,Get-CimClass,Get-CimInstance,Get-CimSession,Get-Clipboard,Get-CmsMessage,Get-Command,Get-ComputerInfo,Get-ComputerRestorePoint,Get-Content,Get-ControlPanelItem,Get-Counter,Get-Credential,Get-Culture,Get-DAPolicyChange,Get-Date,Get-DeliveryOptimizationLog,Get-DeliveryOptimizationPerfSnap,Get-DeliveryOptimizationPerfSnapThisMonth,Get-DeliveryOptimizationStatus,Get-DODownloadMode,Get-DOPercentageMaxBackgroundBandwidth,Get-DOPercentageMaxForegroundBandwidth,Get-Event,Get-EventLog,Get-EventSubscriber,Get-ExecutionPolicy,Get-FormatData,Get-Help,Get-History,Get-Host,Get-HotFix,Get-Item,Get-ItemProperty,Get-ItemPropertyValue,Get-Job,Get-JobTrigger,Get-KdsConfiguration,Get-KdsRootKey,Get-LocalGroup,Get-LocalGroupMember,Get-LocalUser,Get-Location,Get-Member,Get-Module,Get-Package,Get-PackageProvider,Get-PackageSource,Get-PfxCertificate,Get-PfxData,Get-PmemDisk,Get-PmemPhysicalDevice,Get-PmemUnusedRegion,Get-Process,Get-ProcessMitigation,Get-ProvisioningPackage,Get-PSBreakpoint,Get-PSCallStack,Get-PSDrive,Get-PSHostProcessInfo,Get-PSProvider,Get-PSReadlineKeyHandler,Get-PSReadlineOption,Get-PSSession,Get-PSSessionCapability,Get-PSSessionConfiguration,Get-PSSnapin,Get-Random,Get-Runspace,Get-RunspaceDebug,Get-ScheduledJob,Get-ScheduledJobOption,Get-SecureBootPolicy,Get-SecureBootUEFI,Get-Service,Get-TimeZone,Get-TlsCipherSuite,Get-TlsEccCurve,Get-Tpm,Get-TpmEndorsementKeyInfo,Get-TpmSupportedFeature,Get-TraceSource,Get-Transaction,Get-TroubleshootingPack,Get-TrustedProvisioningCertificate,Get-TypeData,Get-UICulture,Get-Unique,Get-Variable,Get-WIMBootEntry,Get-WinAcceptLanguageFromLanguageListOptOut,Get-WinCultureFromLanguageListOptOut,Get-WinDefaultInputMethodOverride,Get-WindowsCapability,Get-WindowsDeveloperLicense,Get-WindowsDriver,Get-WindowsEdition,Get-WindowsErrorReporting,Get-WindowsImage,Get-WindowsImageContent,Get-WindowsOptionalFeature,Get-WindowsPackage,Get-WindowsSearchSetting,Get-WinEvent,Get-WinHomeLocation,Get-WinLanguageBarOption,Get-WinSystemLocale,Get-WinUILanguageOverride,Get-WinUserLanguageList,Get-WmiObject,Get-WSManCredSSP,Get-WSManInstance,Group-Object,Import-Alias,Import-BinaryMiLog,Import-Certificate,Import-Clixml,Import-Counter,Import-Csv,Import-LocalizedData,Import-Module,Import-PackageProvider,Import-PfxCertificate,Import-PSSession,Import-StartLayout,Import-TpmOwnerAuth,Initialize-PmemPhysicalDevice,Initialize-Tpm,Install-Package,Install-PackageProvider,Install-ProvisioningPackage,Install-TrustedProvisioningCertificate,Invoke-CimMethod,Invoke-Command,Invoke-CommandInDesktopPackage,Invoke-DscResource,Invoke-Expression,Invoke-History,Invoke-Item,Invoke-RestMethod,Invoke-TroubleshootingPack,Invoke-WebRequest,Invoke-WmiMethod,Invoke-WSManAction,Join-DtiagnosticResourceManager,Join-Path,Limit-EventLog,Measure-Command,Measure-Object,Mount-AppxVolume,Mount-WindowsImage,Move-AppxPackage,Move-Item,Move-ItemProperty,New-Alias,New-CertificateNotificationTask,New-CimInstance,New-CimSession,New-CimSessionOption,New-DtiagnosticTransaction,New-Event,New-EventLog,New-FileCatalog,New-Item,New-ItemProperty,New-JobTrigger,New-LocalGroup,New-LocalUser,New-Module,New-ModuleManifest,New-NetIPsecAuthProposal,New-NetIPsecMainModeCryptoProposal,New-NetIPsecQuickModeCryptoProposal,New-Object,New-PmemDisk,New-ProvisioningRepro,New-PSDrive,New-PSRoleCapabilityFile,New-PSSession,New-PSSessionConfigurationFile,New-PSSessionOption,New-PSTransportOption,New-PSWorkflowExecutionOption,New-ScheduledJobOption,New-SelfSignedCertificate,New-Service,New-TimeSpan,New-TlsSessionTicketKey,New-Variable,New-WebServiceProxy,New-WindowsCustomImage,New-WindowsImage,New-WinEvent,New-WinUserLanguageList,New-WSManInstance,New-WSManSessionOption,Optimize-AppxProvisionedPackages,Optimize-WindowsImage,Out-Default,Out-File,Out-GridView,Out-Host,Out-Null,Out-Printer,Out-String,Pop-Location,Protect-CmsMessage,Publish-DscConfiguration,Push-Location,Read-Host,Receive-DtiagnosticTransaction,Receive-Job,Receive-PSSession,Register-ArgumentCompleter,Register-CimIndicationEvent,Register-EngineEvent,Register-ObjectEvent,Register-PackageSource,Register-PSSessionConfiguration,Register-ScheduledJob,Register-WmiEvent,Remove-AppxPackage,Remove-AppxProvisionedPackage,Remove-AppxVolume,Remove-BitsTransfer,Remove-CertificateEnrollmentPolicyServer,Remove-CertificateNotificationTask,Remove-CimInstance,Remove-CimSession,Remove-Computer,Remove-Event,Remove-EventLog,Remove-Item,Remove-ItemProperty,Remove-Job,Remove-JobTrigger,Remove-LocalGroup,Remove-LocalGroupMember,Remove-LocalUser,Remove-Module,Remove-PmemDisk,Remove-PSBreakpoint,Remove-PSDrive,Remove-PSReadlineKeyHandler,Remove-PSSession,Remove-PSSnapin,Remove-TypeData,Remove-Variable,Remove-WindowsCapability,Remove-WindowsDriver,Remove-WindowsImage,Remove-WindowsPackage,Remove-WmiObject,Remove-WSManInstance,Rename-Computer,Rename-Item,Rename-ItemProperty,Rename-LocalGroup,Rename-LocalUser,Repair-WindowsImage,Reset-ComputerMachinePassword,Resolve-DnsName,Resolve-Path,Restart-Computer,Restart-Service,Restore-Computer,Resume-BitsTransfer,Resume-Job,Resume-ProvisioningSession,Resume-Service,Save-Help,Save-Package,Save-WindowsImage,Select-Object,Select-String,Select-Xml,Send-DtiagnosticTransaction,Send-MailMessage,Set-Acl,Set-Alias,Set-AppBackgroundTaskResourcePolicy,Set-AppxDefaultVolume,Set-AppXProvisionedDataFile,Set-AuthenticodeSignature,Set-BitsTransfer,Set-CertificateAutoEnrollmentPolicy,Set-CimInstance,Set-Clipboard,Set-Content,Set-Culture,Set-Date,Set-DODownloadMode,Set-DOPercentageMaxBackgroundBandwidth,Set-DOPercentageMaxForegroundBandwidth,Set-DscLocalConfigurationManager,Set-ExecutionPolicy,Set-Item,Set-ItemProperty,Set-JobTrigger,Set-KdsConfiguration,Set-LocalGroup,Set-LocalUser,Set-Location,Set-PackageSource,Set-ProcessMitigation,Set-PSBreakpoint,Set-PSDebug,Set-PSReadlineKeyHandler,Set-PSReadlineOption,Set-PSSessionConfiguration,Set-ScheduledJob,Set-ScheduledJobOption,Set-SecureBootUEFI,Set-Service,Set-StrictMode,Set-TimeZone,Set-TpmOwnerAuth,Set-TraceSource,Set-Variable,Set-WinAcceptLanguageFromLanguageListOptOut,Set-WinCultureFromLanguageListOptOut,Set-WinDefaultInputMethodOverride,Set-WindowsEdition,Set-WindowsProductKey,Set-WindowsSearchSetting,Set-WinHomeLocation,Set-WinLanguageBarOption,Set-WinSystemLocale,Set-WinUILanguageOverride,Set-WinUserLanguageList,Set-WmiInstance,Set-WSManInstance,Set-WSManQuickConfig,Show-Command,Show-ControlPanelItem,Show-EventLog,Show-WindowsDeveloperLicenseRegistration,Sort-Object,Split-Path,Split-WindowsImage,Start-BitsTransfer,Start-DscConfiguration,Start-DtiagnosticResourceManager,Start-Job,Start-Process,Start-Service,Start-Sleep,Start-Transaction,Start-Transcript,Stop-Computer,Stop-DtiagnosticResourceManager,Stop-Job,Stop-Process,Stop-Service,Stop-Transcript,Suspend-BitsTransfer,Suspend-Job,Suspend-Service,Switch-Certificate,Tee-Object,Test-Certificate,Test-ComputerSecureChannel,Test-Connection,Test-DscConfiguration,Test-FileCatalog,Test-KdsRootKey,Test-ModuleManifest,Test-Path,Test-PSSessionConfigurationFile,Test-WSMan,Trace-Command,Unblock-File,Unblock-Tpm,Undo-DtiagnosticTransaction,Undo-Transaction,Uninstall-Package,Uninstall-ProvisioningPackage,Uninstall-TrustedProvisioningCertificate,Unprotect-CmsMessage,Unregister-Event,Unregister-PackageSource,Unregister-PSSessionConfiguration,Unregister-ScheduledJob,Unregister-WindowsDeveloperLicense,Update-FormatData,Update-Help,Update-List,Update-TypeData,Update-WIMBootEntry,Use-Transaction,Use-WindowsUnattend,Wait-Debugger,Wait-Event,Wait-Job,Wait-Process,Where-Object,Write-Debug,Write-Error,Write-EventLog,Write-Host,Write-Information,Write-Output,Write-Progress,Write-Verbose,Write-Warning
	},
	morekeywords={
		Add-BitLockerKeyProtector,Add-DnsClientNrptRule,Add-DtcClusterTMMapping,Add-EtwTraceProvider,Add-InitiatorIdToMaskingSet,Add-MpPreference,Add-NetEventNetworkAdapter,Add-NetEventPacketCaptureProvider,Add-NetEventProvider,Add-NetEventVFPProvider,Add-NetEventVmNetworkAdapter,Add-NetEventVmSwitch,Add-NetEventVmSwitchProvider,Add-NetEventWFPCaptureProvider,Add-NetIPHttpsCertBinding,Add-NetLbfoTeamMember,Add-NetLbfoTeamNic,Add-NetNatExternalAddress,Add-NetNatStaticMapping,Add-NetSwitchTeamMember,Add-Odbsn,Add-PartitionAccessPath,Add-PhysicalDisk,Add-Printer,Add-PrinterDriver,Add-PrinterPort,Add-StorageFaultDomain,Add-TargetPortToMaskingSet,Add-VirtualDiskToMaskingSet,Add-VpnConnection,Add-VpnConnectionRoute,Add-VpnConnectionTriggerApplication,Add-VpnConnectionTriggerDnsConfiguration,Add-VpnConnectionTriggerTrustedNetwork,AfterAll,AfterEach,Assert-MockCalled,Assert-VerifiableMocks,Backup-BitLockerKeyProtector,BackupToAAD-BitLockerKeyProtector,BeforeAll,BeforeEach,Block-FileShareAccess,Block-SmbShareAccess,Clear-BitLockerAutoUnlock,Clear-Disk,Clear-DnsClientCache,Clear-FileStorageTier,Clear-Host,Clear-PcsvDeviceLog,Clear-StorageDiagnosticInfo,Close-SmbOpenFile,Close-SmbSession,Compress-Archive,Configuration,Connect-IscsiTarget,Connect-VirtualDisk,Context,convert,ConvertFrom-SddlString,Copy-NetFirewallRule,Copy-NetIPsecMainModeCryptoSet,Copy-NetIPsecMainModeRule,Copy-NetIPsecPhase1AuthSet,Copy-NetIPsecPhase2AuthSet,Copy-NetIPsecQuickModeCryptoSet,Copy-NetIPsecRule,Debug-FileShare,Debug-MMAppPrelaunch,Debug-StorageSubSystem,Debug-Volume,Describe,Disable-BitLocker,Disable-BitLockerAutoUnlock,Disable-DAManualEntryPointSelection,Disable-Dsebug,Disable-MMAgent,Disable-NetAdapter,Disable-NetAdapterBinding,Disable-NetAdapterChecksumOffload,Disable-NetAdapterEncapsulatedPacketTaskOffload,Disable-NetAdapterIPsecOffload,Disable-NetAdapterLso,Disable-NetAdapterPacketDirect,Disable-NetAdapterPowerManagement,Disable-NetAdapterQos,Disable-NetAdapterRdma,Disable-NetAdapterRsc,Disable-NetAdapterRss,Disable-NetAdapterSriov,Disable-NetAdapterVmq,Disable-NetDnsTransitionConfiguration,Disable-NetFirewallRule,Disable-NetIPHttpsProfile,Disable-NetIPsecMainModeRule,Disable-NetIPsecRule,Disable-NetNatTransitionConfiguration,Disable-NetworkSwitchEthernetPort,Disable-NetworkSwitchFeature,Disable-NetworkSwitchVlan,Disable-OdbcPerfCounter,Disable-PhysicalDiskIdentification,Disable-PnpDevice,Disable-PSTrace,Disable-PSWSManCombinedTrace,Disable-ScheduledTask,Disable-SmbDelegation,Disable-StorageEnclosureIdentification,Disable-StorageEnclosurePower,Disable-StorageHighAvailability,Disable-StorageMaintenanceMode,Disable-WdacBidTrace,Disable-WSManTrace,Disconnect-IscsiTarget,Disconnect-VirtualDisk,Dismount-DiskImage,Enable-BitLocker,Enable-BitLockerAutoUnlock,Enable-DAManualEntryPointSelection,Enable-Dsebug,Enable-MMAgent,Enable-NetAdapter,Enable-NetAdapterBinding,Enable-NetAdapterChecksumOffload,Enable-NetAdapterEncapsulatedPacketTaskOffload,Enable-NetAdapterIPsecOffload,Enable-NetAdapterLso,Enable-NetAdapterPacketDirect,Enable-NetAdapterPowerManagement,Enable-NetAdapterQos,Enable-NetAdapterRdma,Enable-NetAdapterRsc,Enable-NetAdapterRss,Enable-NetAdapterSriov,Enable-NetAdapterVmq,Enable-NetDnsTransitionConfiguration,Enable-NetFirewallRule,Enable-NetIPHttpsProfile,Enable-NetIPsecMainModeRule,Enable-NetIPsecRule,Enable-NetNatTransitionConfiguration,Enable-NetworkSwitchEthernetPort,Enable-NetworkSwitchFeature,Enable-NetworkSwitchVlan,Enable-OdbcPerfCounter,Enable-PhysicalDiskIdentification,Enable-PnpDevice,Enable-PSTrace,Enable-PSWSManCombinedTrace,Enable-ScheduledTask,Enable-SmbDelegation,Enable-StorageEnclosureIdentification,Enable-StorageEnclosurePower,Enable-StorageHighAvailability,Enable-StorageMaintenanceMode,Enable-WdacBidTrace,Enable-WSManTrace,Expand-Archive,Export-ODataEndpointProxy,Export-ScheduledTask,Find-Command,Find-DscResource,Find-Module,Find-NetIPsecRule,Find-NetRoute,Find-RoleCapability,Find-Script,Flush-EtwTraceSession,Format-Hex,Format-Volume,Get-AppBackgroundTask,Get-AppxLastError,Get-AppxLog,Get-AutologgerConfig,Get-BitLockerVolume,Get-ClusteredScheduledTask,Get-DAClientExperienceConfiguration,Get-DAConnectionStatus,Get-DAEntryPointTableItem,Get-DedupProperties,Get-Disk,Get-DiskImage,Get-DiskStorageNodeView,Get-DnsClient,Get-DnsClientCache,Get-DnsClientGlobalSetting,Get-DnsClientNrptGlobal,Get-DnsClientNrptPolicy,Get-DnsClientNrptRule,Get-DnsClientServerAddress,Get-DscConfiguration,Get-DscConfigurationStatus,Get-DscLocalConfigurationManager,Get-DscResource,Get-Dtc,Get-DtcAdvancedHostSetting,Get-DtcAdvancedSetting,Get-DtcClusterDefault,Get-DtcClusterTMMapping,Get-Dtefault,Get-DtcLog,Get-DtcNetworkSetting,Get-DtcTransaction,Get-DtcTransactionsStatistics,Get-DtcTransactionsTraceSession,Get-DtcTransactionsTraceSetting,Get-EtwTraceProvider,Get-EtwTraceSession,Get-FileHash,Get-FileIntegrity,Get-FileShare,Get-FileShareAccessControlEntry,Get-FileStorageTier,Get-InitiatorId,Get-InitiatorPort,Get-InstalledModule,Get-InstalledScript,Get-IscsiConnection,Get-IscsiSession,Get-IscsiTarget,Get-IscsiTargetPortal,Get-IseSnippet,Get-LogProperties,Get-MaskingSet,Get-MMAgent,Get-MockDynamicParameters,Get-MpComputerStatus,Get-MpPreference,Get-MpThreat,Get-MpThreatCatalog,Get-MpThreatDetection,Get-NCSIPolicyConfiguration,Get-Net6to4Configuration,Get-NetAdapter,Get-NetAdapterAdvancedProperty,Get-NetAdapterBinding,Get-NetAdapterChecksumOffload,Get-NetAdapterEncapsulatedPacketTaskOffload,Get-NetAdapterHardwareInfo,Get-NetAdapterIPsecOffload,Get-NetAdapterLso,Get-NetAdapterPacketDirect,Get-NetAdapterPowerManagement,Get-NetAdapterQos,Get-NetAdapterRdma,Get-NetAdapterRsc,Get-NetAdapterRss,Get-NetAdapterSriov,Get-NetAdapterSriovVf,Get-NetAdapterStatistics,Get-NetAdapterVmq,Get-NetAdapterVMQQueue,Get-NetAdapterVPort,Get-NetCompartment,Get-NetConnectionProfile,Get-NetDnsTransitionConfiguration,Get-NetDnsTransitionMonitoring,Get-NetEventNetworkAdapter,Get-NetEventPacketCaptureProvider,Get-NetEventProvider,Get-NetEventSession,Get-NetEventVFPProvider,Get-NetEventVmNetworkAdapter,Get-NetEventVmSwitch,Get-NetEventVmSwitchProvider,Get-NetEventWFPCaptureProvider,Get-NetFirewallAddressFilter,Get-NetFirewallApplicationFilter,Get-NetFirewallInterfaceFilter,Get-NetFirewallInterfaceTypeFilter,Get-NetFirewallPortFilter,Get-NetFirewallProfile,Get-NetFirewallRule,Get-NetFirewallSecurityFilter,Get-NetFirewallServiceFilter,Get-NetFirewallSetting,Get-NetIPAddress,Get-NetIPConfiguration,Get-NetIPHttpsConfiguration,Get-NetIPHttpsState,Get-NetIPInterface,Get-NetIPseospSetting,Get-NetIPsecMainModeCryptoSet,Get-NetIPsecMainModeRule,Get-NetIPsecMainModeSA,Get-NetIPsecPhase1AuthSet,Get-NetIPsecPhase2AuthSet,Get-NetIPsecQuickModeCryptoSet,Get-NetIPsecQuickModeSA,Get-NetIPsecRule,Get-NetIPv4Protocol,Get-NetIPv6Protocol,Get-NetIsatapConfiguration,Get-NetLbfoTeam,Get-NetLbfoTeamMember,Get-NetLbfoTeamNic,Get-NetNat,Get-NetNatExternalAddress,Get-NetNatGlobal,Get-NetNatSession,Get-NetNatStaticMapping,Get-NetNatTransitionConfiguration,Get-NetNatTransitionMonitoring,Get-NetNeighbor,Get-NetOffloadGlobalSetting,Get-NetPrefixPolicy,Get-NetQosPolicy,Get-NetRoute,Get-NetSwitchTeam,Get-NetSwitchTeamMember,Get-NetTCPConnection,Get-NetTCPSetting,Get-NetTeredoConfiguration,Get-NetTeredoState,Get-NetTransportFilter,Get-NetUDPEndpoint,Get-NetUDPSetting,Get-NetworkSwitchEthernetPort,Get-NetworkSwitchFeature,Get-NetworkSwitchGlobalData,Get-NetworkSwitchVlan,Get-Odbriver,Get-Odbsn,Get-OdbcPerfCounter,Get-OffloadDataTransferSetting,Get-OperationValidation,Get-Partition,Get-PartitionSupportedSize,Get-PcsvDevice,Get-PcsvDeviceLog,Get-PhysicalDisk,Get-PhysicalDiskStorageNodeView,Get-PhysicalExtent,Get-PhysicalExtentAssociation,Get-PnpDevice,Get-PnpDeviceProperty,Get-PrintConfiguration,Get-Printer,Get-PrinterDriver,Get-PrinterPort,Get-PrinterProperty,Get-PrintJob,Get-PSRepository,Get-ResiliencySetting,Get-ScheduledTask,Get-ScheduledTaskInfo,Get-SmbBandWidthLimit,Get-SmbClientConfiguration,Get-SmbClientNetworkInterface,Get-SmbConnection,Get-SmbDelegation,Get-SmbGlobalMapping,Get-SmbMapping,Get-SmbMultichannelConnection,Get-SmbMultichannelConstraint,Get-SmbOpenFile,Get-SmbServerConfiguration,Get-SmbServerNetworkInterface,Get-SmbSession,Get-SmbShare,Get-SmbShareAccess,Get-SmbWitnessClient,Get-StartApps,Get-StorageAdvancedProperty,Get-StorageDiagnosticInfo,Get-StorageEnclosure,Get-StorageEnclosureStorageNodeView,Get-StorageEnclosureVendorData,Get-StorageExtendedStatus,Get-StorageFaultDomain,Get-StorageFileServer,Get-StorageFirmwareInformation,Get-StorageHealthAction,Get-StorageHealthReport,Get-StorageHealthSetting,Get-StorageJob,Get-StorageNode,Get-StoragePool,Get-StorageProvider,Get-StorageReliabilityCounter,Get-StorageSetting,Get-StorageSubSystem,Get-StorageTier,Get-StorageTierSupportedSize,Get-SupportedClusterSizes,Get-SupportedFileSystems,Get-TargetPort,Get-TargetPortal,Get-TestDriveItem,Get-Verb,Get-VirtualDisk,Get-VirtualDiskSupportedSize,Get-Volume,Get-VolumeCorruptionCount,Get-VolumeScrubPolicy,Get-VpnConnection,Get-VpnConnectionTrigger,Get-WdacBidTrace,Get-WindowsUpdateLog,Get-WUAVersion,Get-WUIsPendingReboot,Get-WULastInstallationDate,Get-WULastScanSuccessDate,Grant-FileShareAccess,Grant-SmbShareAccess,help,Hide-VirtualDisk,Import-IseSnippet,Import-PowerShellDataFile,ImportSystemModules,In,Initialize-Disk,InModuleScope,Install-Dtc,Install-Module,Install-Script,Install-WUUpdates,Invoke-AsWorkflow,Invoke-Mock,Invoke-OperationValidation,Invoke-Pester,It,Lock-BitLocker,mkdir,Mock,more,Mount-DiskImage,Move-SmbWitnessClient,New-AutologgerConfig,New-DAEntryPointTableItem,New-DscChecksum,New-EapConfiguration,New-EtwTraceSession,New-FileShare,New-Fixture,New-Guid,New-IscsiTargetPortal,New-IseSnippet,New-MaskingSet,New-NetAdapterAdvancedProperty,New-NetEventSession,New-NetFirewallRule,New-NetIPAddress,New-NetIPHttpsConfiguration,New-NetIPseospSetting,New-NetIPsecMainModeCryptoSet,New-NetIPsecMainModeRule,New-NetIPsecPhase1AuthSet,New-NetIPsecPhase2AuthSet,New-NetIPsecQuickModeCryptoSet,New-NetIPsecRule,New-NetLbfoTeam,New-NetNat,New-NetNatTransitionConfiguration,New-NetNeighbor,New-NetQosPolicy,New-NetRoute,New-NetSwitchTeam,New-NetTransportFilter,New-NetworkSwitchVlan,New-Partition,New-PesterOption,New-PSWorkflowSession,New-ScheduledTask,New-ScheduledTaskAction,New-ScheduledTaskPrincipal,New-ScheduledTaskSettingsSet,New-ScheduledTaskTrigger,New-ScriptFileInfo,New-SmbGlobalMapping,New-SmbMapping,New-SmbMultichannelConstraint,New-SmbShare,New-StorageFileServer,New-StoragePool,New-StorageSubsystemVirtualDisk,New-StorageTier,New-TemporaryFile,New-VirtualDisk,New-VirtualDiskClone,New-VirtualDiskSnapshot,New-Volume,New-VpnServerAddress,Open-NetGPO,Optimize-StoragePool,Optimize-Volume,oss,Pause,prompt,PSConsoleHostReadline,Publish-Module,Publish-Script,Read-PrinterNfcTag,Register-ClusteredScheduledTask,Register-DnsClient,Register-IscsiSession,Register-PSRepository,Register-ScheduledTask,Register-StorageSubsystem,Remove-AutologgerConfig,Remove-BitLockerKeyProtector,Remove-DAEntryPointTableItem,Remove-DnsClientNrptRule,Remove-DscConfigurationDocument,Remove-DtcClusterTMMapping,Remove-EtwTraceProvider,Remove-FileShare,Remove-InitiatorId,Remove-InitiatorIdFromMaskingSet,Remove-IscsiTargetPortal,Remove-MaskingSet,Remove-MpPreference,Remove-MpThreat,Remove-NetAdapterAdvancedProperty,Remove-NetEventNetworkAdapter,Remove-NetEventPacketCaptureProvider,Remove-NetEventProvider,Remove-NetEventSession,Remove-NetEventVFPProvider,Remove-NetEventVmNetworkAdapter,Remove-NetEventVmSwitch,Remove-NetEventVmSwitchProvider,Remove-NetEventWFPCaptureProvider,Remove-NetFirewallRule,Remove-NetIPAddress,Remove-NetIPHttpsCertBinding,Remove-NetIPHttpsConfiguration,Remove-NetIPseospSetting,Remove-NetIPsecMainModeCryptoSet,Remove-NetIPsecMainModeRule,Remove-NetIPsecMainModeSA,Remove-NetIPsecPhase1AuthSet,Remove-NetIPsecPhase2AuthSet,Remove-NetIPsecQuickModeCryptoSet,Remove-NetIPsecQuickModeSA,Remove-NetIPsecRule,Remove-NetLbfoTeam,Remove-NetLbfoTeamMember,Remove-NetLbfoTeamNic,Remove-NetNat,Remove-NetNatExternalAddress,Remove-NetNatStaticMapping,Remove-NetNatTransitionConfiguration,Remove-NetNeighbor,Remove-NetQosPolicy,Remove-NetRoute,Remove-NetSwitchTeam,Remove-NetSwitchTeamMember,Remove-NetTransportFilter,Remove-NetworkSwitchEthernetPortIPAddress,Remove-NetworkSwitchVlan,Remove-Odbsn,Remove-Partition,Remove-PartitionAccessPath,Remove-PhysicalDisk,Remove-Printer,Remove-PrinterDriver,Remove-PrinterPort,Remove-PrintJob,Remove-SmbBandwidthLimit,Remove-SmbGlobalMapping,Remove-SmbMapping,Remove-SmbMultichannelConstraint,Remove-SmbShare,Remove-StorageFaultDomain,Remove-StorageFileServer,Remove-StorageHealthIntent,Remove-StorageHealthSetting,Remove-StoragePool,Remove-StorageTier,Remove-TargetPortFromMaskingSet,Remove-VirtualDisk,Remove-VirtualDiskFromMaskingSet,Remove-VpnConnection,Remove-VpnConnectionRoute,Remove-VpnConnectionTriggerApplication,Remove-VpnConnectionTriggerDnsConfiguration,Remove-VpnConnectionTriggerTrustedNetwork,Rename-DAEntryPointTableItem,Rename-MaskingSet,Rename-NetAdapter,Rename-NetFirewallRule,Rename-NetIPHttpsConfiguration,Rename-NetIPsecMainModeCryptoSet,Rename-NetIPsecMainModeRule,Rename-NetIPsecPhase1AuthSet,Rename-NetIPsecPhase2AuthSet,Rename-NetIPsecQuickModeCryptoSet,Rename-NetIPsecRule,Rename-NetLbfoTeam,Rename-NetSwitchTeam,Rename-Printer,Repair-FileIntegrity,Repair-VirtualDisk,Repair-Volume,Reset-DAClientExperienceConfiguration,Reset-DAEntryPointTableItem,Reset-DtcLog,Reset-NCSIPolicyConfiguration,Reset-Net6to4Configuration,Reset-NetAdapterAdvancedProperty,Reset-NetDnsTransitionConfiguration,Reset-NetIPHttpsConfiguration,Reset-NetIsatapConfiguration,Reset-NetTeredoConfiguration,Reset-PhysicalDisk,Reset-StorageReliabilityCounter,Resize-Partition,Resize-StorageTier,Resize-VirtualDisk,Restart-NetAdapter,Restart-PcsvDevice,Restart-PrintJob,Restore-DscConfiguration,Restore-NetworkSwitchConfiguration,Resume-BitLocker,Resume-PrintJob,Revoke-FileShareAccess,Revoke-SmbShareAccess,SafeGetCommand,Save-EtwTraceSession,Save-Module,Save-NetGPO,Save-NetworkSwitchConfiguration,Save-Script,Send-EtwTraceSession,Set-AutologgerConfig,Set-ClusteredScheduledTask,Set-DAClientExperienceConfiguration,Set-DAEntryPointTableItem,Set-Disk,Set-DnsClient,Set-DnsClientGlobalSetting,Set-DnsClientNrptGlobal,Set-DnsClientNrptRule,Set-DnsClientServerAddress,Set-DtcAdvancedHostSetting,Set-DtcAdvancedSetting,Set-DtcClusterDefault,Set-DtcClusterTMMapping,Set-Dtefault,Set-DtcLog,Set-DtcNetworkSetting,Set-DtcTransaction,Set-DtcTransactionsTraceSession,Set-DtcTransactionsTraceSetting,Set-DynamicParameterVariables,Set-EtwTraceProvider,Set-FileIntegrity,Set-FileShare,Set-FileStorageTier,Set-InitiatorPort,Set-IscsiChapSecret,Set-LogProperties,Set-MMAgent,Set-MpPreference,Set-NCSIPolicyConfiguration,Set-Net6to4Configuration,Set-NetAdapter,Set-NetAdapterAdvancedProperty,Set-NetAdapterBinding,Set-NetAdapterChecksumOffload,Set-NetAdapterEncapsulatedPacketTaskOffload,Set-NetAdapterIPsecOffload,Set-NetAdapterLso,Set-NetAdapterPacketDirect,Set-NetAdapterPowerManagement,Set-NetAdapterQos,Set-NetAdapterRdma,Set-NetAdapterRsc,Set-NetAdapterRss,Set-NetAdapterSriov,Set-NetAdapterVmq,Set-NetConnectionProfile,Set-NetDnsTransitionConfiguration,Set-NetEventPacketCaptureProvider,Set-NetEventProvider,Set-NetEventSession,Set-NetEventVFPProvider,Set-NetEventVmSwitchProvider,Set-NetEventWFPCaptureProvider,Set-NetFirewallAddressFilter,Set-NetFirewallApplicationFilter,Set-NetFirewallInterfaceFilter,Set-NetFirewallInterfaceTypeFilter,Set-NetFirewallPortFilter,Set-NetFirewallProfile,Set-NetFirewallRule,Set-NetFirewallSecurityFilter,Set-NetFirewallServiceFilter,Set-NetFirewallSetting,Set-NetIPAddress,Set-NetIPHttpsConfiguration,Set-NetIPInterface,Set-NetIPseospSetting,Set-NetIPsecMainModeCryptoSet,Set-NetIPsecMainModeRule,Set-NetIPsecPhase1AuthSet,Set-NetIPsecPhase2AuthSet,Set-NetIPsecQuickModeCryptoSet,Set-NetIPsecRule,Set-NetIPv4Protocol,Set-NetIPv6Protocol,Set-NetIsatapConfiguration,Set-NetLbfoTeam,Set-NetLbfoTeamMember,Set-NetLbfoTeamNic,Set-NetNat,Set-NetNatGlobal,Set-NetNatTransitionConfiguration,Set-NetNeighbor,Set-NetOffloadGlobalSetting,Set-NetQosPolicy,Set-NetRoute,Set-NetTCPSetting,Set-NetTeredoConfiguration,Set-NetUDPSetting,Set-NetworkSwitchEthernetPortIPAddress,Set-NetworkSwitchPortMode,Set-NetworkSwitchPortProperty,Set-NetworkSwitchVlanProperty,Set-Odbriver,Set-Odbsn,Set-Partition,Set-PcsvDeviceBootConfiguration,Set-PcsvDeviceNetworkConfiguration,Set-PcsvDeviceUserPassword,Set-PhysicalDisk,Set-PrintConfiguration,Set-Printer,Set-PrinterProperty,Set-PSRepository,Set-ResiliencySetting,Set-ScheduledTask,Set-SmbBandwidthLimit,Set-SmbClientConfiguration,Set-SmbPathAcl,Set-SmbServerConfiguration,Set-SmbShare,Set-StorageFileServer,Set-StorageHealthSetting,Set-StoragePool,Set-StorageProvider,Set-StorageSetting,Set-StorageSubSystem,Set-StorageTier,Set-TestInconclusive,Setup,Set-VirtualDisk,Set-Volume,Set-VolumeScrubPolicy,Set-VpnConnection,Set-VpnConnectionIPsecConfiguration,Set-VpnConnectionProxy,Set-VpnConnectionTriggerDnsConfiguration,Set-VpnConnectionTriggerTrustedNetwork,Should,Show-NetFirewallRule,Show-NetIPsecRule,Show-VirtualDisk,Start-AppBackgroundTask,Start-AutologgerConfig,Start-Dtc,Start-DtcTransactionsTraceSession,Start-EtwTraceSession,Start-MpScan,Start-MpWDOScan,Start-NetEventSession,Start-PcsvDevice,Start-ScheduledTask,Start-StorageDiagnosticLog,Start-Trace,Start-WUScan,Stop-DscConfiguration,Stop-Dtc,Stop-DtcTransactionsTraceSession,Stop-EtwTraceSession,Stop-NetEventSession,Stop-PcsvDevice,Stop-ScheduledTask,Stop-StorageDiagnosticLog,Stop-StorageJob,Stop-Trace,Suspend-BitLocker,Suspend-PrintJob,Sync-NetIPsecRule,TabExpansion2,Test-Dtc,Test-NetConnection,Test-ScriptFileInfo,Unblock-FileShareAccess,Unblock-SmbShareAccess,Uninstall-Dtc,Uninstall-Module,Uninstall-Script,Unlock-BitLocker,Unregister-AppBackgroundTask,Unregister-ClusteredScheduledTask,Unregister-IscsiSession,Unregister-PSRepository,Unregister-ScheduledTask,Unregister-StorageSubsystem,Update-Disk,Update-DscConfiguration,Update-EtwTraceSession,Update-HostStorageCache,Update-IscsiTarget,Update-IscsiTargetPortal,Update-Module,Update-ModuleManifest,Update-MpSignature,Update-NetIPsecRule,Update-Script,Update-ScriptFileInfo,Update-SmbMultichannelConnection,Update-StorageFirmware,Update-StoragePool,Update-StorageProviderCache,Write-DtcTransactionsTraceSession,Write-PrinterNfcTag,Write-VolumeCache
	},
	morekeywords={Do,Else,For,ForEach,Function,If,In,Until,While},
	alsodigit={-},
	sensitive=false,
	morecomment=[l]{\#},
	morecomment=[n]{<\#}{\#>},
	morestring=[b]{"},
	morestring=[b]{'},
	morestring=[s]{@'}{'@},
	morestring=[s]{@"}{"@}
}
\definecolor{dkgreen}{rgb}{0,0.6,0}
\definecolor{mauve}{rgb}{0.88, 0.69, 1.0}
\lstdefinelanguage{batch}{
    morekeywords={not,exist,mkdir,echo,pause},
    morekeywords={Do,Else,For,ForEach,Function,If,In,Until,While},
    alsodigit={-},
    sensitive=false,
    morecomment=[l]{\#},
    morecomment=[n]{<\#}{\#>},
    morestring=[b]{"},
    morestring=[b]{'},
    morestring=[s]{@'}{'@},
    morestring=[s]{@"}{"@},
    numbers=left,
    numberstyle=\tiny,
    basicstyle={\small\ttfamily},
    stringstyle=\color{dkgreen},
    keywordstyle=\color{blue},
    frame=tb,
}

\makeatletter

\newcommand*{\overrightharpoonup}{\mathpalette{\overarrow@\rightharpoonupfill@}}
\newcommand*{\rightharpoonupfill@}{\arrowfill@\relbar\relbar\rightharpoonup}
\makeatother

\makeatletter
\newcommand*{\overleftharpoonup}{\mathpalette{\overarrow@\leftharpoonupfill@}}
\newcommand*{\leftharpoonupfill@}{\arrowfill@\relbar\relbar\leftharpoonup}
\makeatother

\definecolor{gray90}{gray}{0.9}

\makeatletter
\newcommand*\bigcdot{\mathpalette\bigcdot@{1}}
\newcommand*\bigcdot@[2]{\mathbin{\vcenter{\hbox{\scalebox{#2}{$\m@th#1\bullet$}}}}}
\makeatother

\crefname{section}{Section}{Sections}
\crefname{equation}{Equation}{Equations}
\crefname{figure}{Figure}{Figures}
\crefname{table}{Table}{Tables}
\crefname{listing}{Listing}{Listings}

\setlist[itemize]{leftmargin=10pt,itemindent=0pt,topsep=2pt,partopsep=2.5pt,parsep=1pt,itemsep=1.5pt,listparindent=\parindent{}}
\setlist[enumerate]{leftmargin=16pt,itemindent=0pt,topsep=2pt,partopsep=2.5pt,parsep=1pt,itemsep=1.5pt,listparindent=\parindent{}}

\title{Performance Analysis of Hardware-Accelerated 10-Bit 4:2:2 Encoding with Split-Frame Encoding for High-Fidelity V-PCC Streaming}
\name{Kasidis Arunruangsirilert, Jiro Katto}
\address{School of Fundamental Science and Engineering, Waseda University, Tokyo, Japan}

\begin{document}
\bstctlcite{IEEEexample:BSTcontrol}
%
\maketitle

\setstretch{0.96}
\begin{abstract}
\setstretch{0.88}

Video-based Point Cloud Compression (V-PCC) encodes volumetric data by projecting 3D geometry and texture onto 2D video frames. To prevent spatial distortion and color bleeding during 3D reconstruction, this process requires 10-bit color depth and 4:2:2 chroma subsampling, rather than the standard 8-bit 4:2:0 format. Additionally, capturing high-density dynamic point clouds requires demanding encoding parameters, such as 8K resolution at framerates up to 120 fps. Historically, the lack of 4:2:2 chroma support in older GPU hardware encoders restricted real-time V-PCC to custom Application-Specific Integrated Circuits (ASICs). However, the recent introduction of NVIDIA's Blackwell GPU architecture, featuring on-chip hardware encoders with 10-bit 4:2:2 support, presents an opportunity to shift this workload to general-purpose hardware. This paper investigates the feasibility of such an approach. Using a commercially available Blackwell GPU equipped with four parallel on-die hardware encoders as a testbed, we evaluate the throughput, rate-distortion (RD) performance, and power consumption of 8K 10-bit 4:2:2 HEVC across various Split-Frame Encoding (SFE) configurations. Our results demonstrate that 4-way SFE achieves an encoding throughput of 122 fps, successfully meeting the strict real-time constraints of high-density V-PCC. Although the inability to exploit spatial redundancies across slice boundaries results in a BD-Rate penalty of up to 5\%, the measured throughput and power efficiency establish standard, commercial off-the-shelf GPUs as a highly viable baseline for real-time volumetric video streaming.

\end{abstract}
\setstretch{0.91}
\vspace{-3mm}
\section{Introduction}
\label{sec:intro}
\vspace{-3mm}

\blfootnote{This paper is supported by the Ministry of Internal Affairs and Communications (MIC) Project for Efficient Frequency Utilization Toward Wireless IP Multicasting and the Japan Science and Technology Agency (JST) CRONOS Grant Number JPMJCS25N2.}

\vspace{-2mm}

Volumetric media formats, such as point clouds, require significant data compression for storage and transmission \cite{10.1145/3682062}. V-PCC addresses this by packing 3D spatial geometry and surface attributes into 2D frames, which are then compressed using conventional 2D video codecs such as HEVC, AV1, or VVC. In this paradigm, the structural accuracy of the reconstructed 3D model is entirely dependent on the compression efficiency and pixel-level fidelity of the 2D video streams \cite{10.1145/3690641}. 

One of the key constraints in V-PCC is chroma subsampling, which dictates the encoding resolution of chroma components relative to the luminance counterpart. In standard video delivery, 8-bit color depth and 4:2:0 chroma subsampling are deemed sufficient due to the limitations of human visual perception. However, using 4:2:0 subsampling to encode the V-PCC video channels (attribute, geometry, and occupancy) will inevitably introduce interpolation errors \cite{11396839}. When re-projected into 3D space, these errors manifest as visual artifacts such as geometric deformations, surface tearing, and color bleeding \cite{10470357}. Furthermore, using standard 8-bit video, which is limited to 256 discrete values, severely restricts the precision needed for accurate 3D rendering. Because V-PCC encodes spatial coordinates as depth maps, 8-bit quantization restricts spatial resolution, resulting in geometric jitter, jagged edges, and visible surface holes. Additionally, packing color attributes into 8-bit 2D atlases causes muddy textures and color banding \cite{10682566}. Consequently, preserving point cloud integrity strictly requires encoders operating at high bit depths and high chroma fidelity, typically utilizing 10-bit 4:2:2 or 4:4:4 formats \cite{10403987}. Moreover, as point cloud density and temporal resolution scale to support realistic capture, the encoding pipeline must handle 8K (7680×4320) resolution at a framerate of 120 frames per second (fps).

\begin{figure}[t!]
\centering\includesvg[width=0.92\linewidth,inkscapelatex=false]{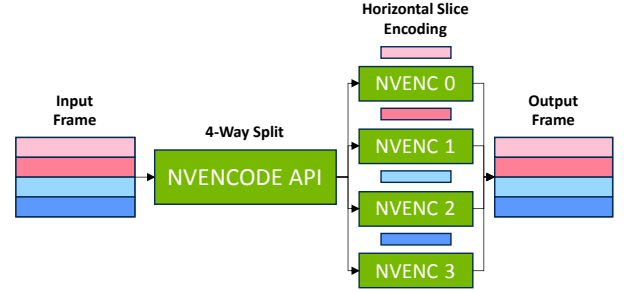}
\vspace{-3mm}
\caption{Pipeline of 4-Way Split-Frame Encoding (SFE)}
\label{fig:Pipeline}
\vspace{-3.5mm}
\end{figure}

\begin{figure}[t!]
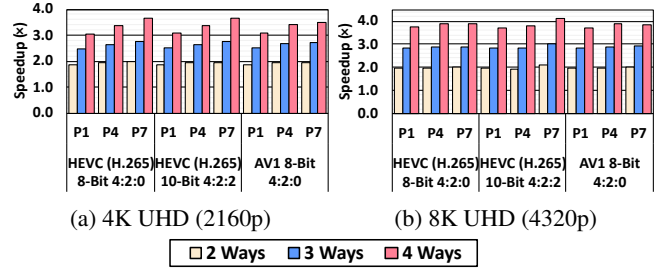

\centering
\begin{subfigure}[t]{.49\linewidth}
  \centering\includesvg[width=0.99\linewidth,inkscapelatex=false]{4KSpeedUp.svg}
  \vspace{-4.5mm}
  \caption{4K UHD (2160p)}
  \label{fig:4KSpeedUp}
\end{subfigure}
\begin{subfigure}[t]{.49\linewidth}
\centering\includesvg[width=0.99\linewidth,inkscapelatex=false]{8KSpeedUp.svg}
  \vspace{-4.5mm}
  \caption{8K UHD (4320p)}
  \label{fig:8KSpeedUp}
\end{subfigure}\\
\begin{subfigure}[t]{.95\linewidth}
\vspace{0.2mm}
  \centering\includesvg[width=0.5\linewidth,inkscapelatex=false]{Legend.svg}
  \vspace{-3.5mm}
\end{subfigure}
\setlength{\belowcaptionskip}{-18pt}
  \vspace{1mm}
\caption{Speedup Scaling of each SFE configuration relative to SFE disabled for HEVC (H.265) 10-Bit 4:2:2 encoding.}
\label{fig:SpeedUp}
\vspace{-7mm}
\end{figure}

The combination of 8K resolution, 120 fps throughput, and 10-bit 4:2:2 subsampling creates a computational workload that exceeds the capabilities of CPU-based software encoders. Until recently, real-time encoding of such video streams was the exclusive domain of specialized Application-Specific Integrated Circuits (ASICs) or multi-FPGA arrays \cite{8388869, 9013947, 10906555}. The lack of standard general-purpose hardware support has introduced significant system-level friction for deploying real-time V-PCC streaming architectures, restricting high-fidelity research and deployment to well-funded entities. \looseness=-1

Recently, the landscape of video encoding has been shifting with the advent of high-performance General-Purpose Graphics Processing Units (GPGPUs). One of the major GPU manufacturers, NVIDIA, has integrated robust hardware video encoders (NVENC) directly onto the GPU die. While earlier architectures were often limited to standard 8-bit 4:2:0 consumer formats, recent silicon advancements have bridged this capability gap. Notably, NVIDIA’s latest Blackwell architecture introduces hardware-accelerated 10-bit 4:2:2 encoding to commercial off-the-shelf (COTS) hardware \cite{nvidia_bw_2024}. Furthermore, to tackle the extreme pixel rate of modern video encoding tasks, the flagship RTX PRO 6000 Blackwell GPU, based on the GB202 die, integrates four independent NVENC engines, which enable a 4-Way Split-Frame Encoding (SFE) configuration \cite{nvidia_2024b}. SFE partitions a single frame into independent horizontal slices, assigns them to individual NVENC engines, and then merges the bitstreams (see \cref{fig:Pipeline}), a strategy mirroring the slice division method originally proposed by Iwasaki et al. for broadcasting standards \cite{9043021}. While recent evaluations on the previous Ada Lovelace architecture demonstrated that 2-way SFE could hit an 8K 60fps target in 8-bit 4:2:0 format with negligible quality loss \cite{11417632, nvidia_2024}, the four-engine design in Blackwell provides the necessary parallel scaling to achieve real-time 8K 120fps processing in 10-bit 4:2:2 format on a single GPU. \looseness=-1

This paper presents an empirical evaluation of hardware-accelerated 10-bit 4:2:2 HEVC encoding for volumetric video applications. By quantifying throughput, rate-distortion (RD) performance, and power consumption for 4K and 8K encoding in HEVC and AV1 codecs across various SFE configurations, this study establishes baseline metrics for utilizing COTS hardware to support real-time V-PCC pipelines.

\vspace{-3mm}
\section{Experiment Setup}
\vspace{-1mm}
\subsection{Encoding Throughput Benchmark}

To evaluate the upper bounds of the hardware video encoder on modern COTS GPUs, we utilized the NVIDIA RTX PRO 6000 Blackwell GPU as our testbed. This GPU was selected because it integrates four NVENC engines, the maximum count in the architecture, permitting a 4-Way SFE configuration. While the tests were conducted on this SKU, the results are generalizable to the broader Blackwell architecture GPUs and similar multi-encoder topologies. To measure the peak encoding throughput of these NVENC engines and eliminate system-level bottlenecks, we adopted a synthetic frame generation methodology. Benchmarking peak throughput using real 8K video is technically infeasible: hardware decoders (NVDEC) are slower than the fastest encoder preset (P1), and using a software decoder hits the PCIe 5.0 x16 transfer limit (uncompressed 8K 120fps 10-bit 4:2:2 saturates the 64 GB/s directional bus). Furthermore, unlike software encoders or certain Video Processing Units (VPUs, e.g., NETINT), where varying macroblock complexity impacts encoding throughput, fixed-block GPU encoders process at a constant rate regardless of scene complexity. Consequently, generating frames directly in VRAM is the only scientifically valid method to isolate and measure the NVENC silicon's true limits. Following this peer-reviewed and validated protocol \cite{11417632}, we utilized FFmpeg's \textit{lavfi} source to generate a black frame, which was immediately uploaded to the GPU VRAM using the \textit{hwupload} filter and looped internally for the duration of the encoding session.


We evaluated the encoding throughput of three distinct codec and format configurations representing consumer, broadcast, and streaming standards: HEVC (H.265) 8-bit 4:2:0 (Baseline), HEVC (H.265) 10-bit 4:2:2, which serves as the primary focus for V-PCC encoding in this study, and AV1 8-bit 4:2:0. It should be noted that Blackwell GPUs do not support 10-bit 4:2:2 encoding for the AV1 codec; the results here are provided for reference. The evaluation covered all available SFE configurations supported by the Blackwell architecture: 1-way (single encoder), 2-way, 3-way, and 4-way split modes. For each mode, we tested across three performance presets: P1 (Fastest), P4 (Medium), and P7 (Slowest/High Quality), combined with \textit{Low Latency (LL)} and \textit{High Quality (HQ)} tuning. To ensure statistical stability, the 8K UHD (7680×4320) tests were conducted at a constant bitrate (CBR) of 200 Mbps over a duration of 3,600 frames. Conversely, 4K UHD (3840×2160) tests were conducted at 50 Mbps over 14,400 frames. The average encoding speed (fps) was recorded for each trial. The specific encoding scripts used are provided in the \textit{Supplementary Material}.
\vspace{-5mm}

\begin{table}[!tbp]
\vspace{-2.5mm}
\setstretch{0.82}
\caption{Hardware and Software Configuration}
\centering
\label{tab:hardware}
\resizebox{8.5cm}{!}{\begin{tabular}{@{}ll@{}}
\toprule
\multicolumn{2}{c}{Encoding System}\\
\midrule
Hardware                 & Description  \\\midrule
CPU & Intel(R) Core(TM) Ultra 9 285K (OC to 5.4 GHz) \\
RAM & Dual-Channel DDR5 128 GB (4×32 GB) @ 6000 MHz \\ 
GPU & NVIDIA RTX PRO 6000 Blackwell Max-Q Workstation \\\midrule
Software & Version \\\midrule
OS & Microsoft Windows Server 2025 Datacenter Build 26100\\
ffmpeg & N-122268-g0dfaed77a6 Patched with PR \#21371\\
NVIDIA GPU Driver & NVIDIA RTX Driver Release 580 R580 U5 (581.80)\\\midrule
\multicolumn{2}{c}{VMAF Calculation System}\\
\midrule
Hardware                 & Description  \\\midrule
CPU & Intel(R) Core(TM) Ultra 9 285K \\
RAM & Dual-Channel DDR5 128 GB (4×32 GB) @ 4400 MHz \\ 
GPU & NVIDIA RTX PRO 5000 48GB Blackwell\\\midrule
Software & Version \\\midrule
OS & Ubuntu 24.04.3 LTS\\
ffmpeg & N-122271-g0629780cf6\\
libvmaf & v3.0.0 (b9ac69e6)\\
VMAF Model & vmaf\_4k\_v0.6.1neg \\
NVIDIA GPU Driver & 580.126.09 \\
NVIDIA CUDA Compiler & cuda\_13.1.r13.1\/compiler.36836380\_0 \\
\bottomrule
\end{tabular}}
\vspace{-6.5mm}
\end{table}
\vspace{1mm}

\subsubsection{Test Datasets}
\vspace{-1mm}
Although this study targets V-PCC, evaluating projected frames across our extensive preset and bitrate matrix is computationally infeasible. Therefore, we utilized standard 2D video datasets, as the VMAF/PSNR calculations can be CUDA-accelerated, enabling large-scale evaluation and direct comparison to existing literature \cite{10637525, 11396901, 11417632, evolutionnvencefficiencylongitudinal}. To evaluate compression efficiency across a diverse range of content, we utilized five distinct datasets comprising both broadcast standard sequences and high-fidelity streaming content. Except for the ITE dataset, which features Standard Dynamic Range (SDR) with \textit{BT2020} Wide Color Gamut (WCG), the rest of the datasets are in High Dynamic Range (HDR) with \textit{SMPTE432} (Display P3) Color Space. The list of all datasets is as follows:
\begin{enumerate}
  \item \textbf{ITE Ultra-High Definition Standard Test Sequences (Series A) \cite{ITE_2016}:} As described in the official documentation, this dataset includes 10 sequences at 4K resolution and 11 sequences at 8K resolution. All sequences have a framerate of 59.94 fps and are 15 seconds in length.
  \item \textbf{Netflix Sol Levante  \cite{netflix}:} A 4K (3840×2160) animation sequence at 24 fps with a running time of 4:32 minutes.
  \item \textbf{Netflix Meridian  \cite{netflix}:} A 4K (3840×2160) live-action sequence at 59.94 fps with a running time of 11:58 minutes.
  \item \textbf{Netflix Nocturne  \cite{netflix}:} A 4K (3840×2160) High-Frame-Rate (HFR) sequence at 120 fps with a running time of 11:04 minutes.
  \item \textbf{Netflix Chimera  \cite{netflix}:} A DCI 4K (4096×2160) sequence at 59.94 fps consisting of 23 distinct scenes with a total runtime of 30:49 minutes.
\end{enumerate}

\vspace{-4.5mm}

\subsubsection{Mezzanine File Preparation}
\vspace{-1mm}

The source content was originally provided in uncompressed RAW formats (TIFF, EXR, DPX). To facilitate efficient large-scale testing and eliminate disk I/O bottlenecks during the encoding trials, we generated high-fidelity mezzanine files. These intermediates permitted the utilization of the on-die hardware decoder (NVDEC) over the software counterpart.

The mezzanine files were encoded using the \textit{x265} software encoder with a Constant Rate Factor (CRF) of 10 and the \textit{medium} preset to ensure mathematically near-lossless fidelity. We then encoded all of the sequences using the \textit{yuv422p10le} pixel format. We utilized the \textit{bt2020nc} color matrix with \textit{SMPTE432} (Display P3) color primaries. For HDR content (e.g., Meridian), the transfer characteristics were set to SMPTE 2084 (PQ), while SDR WCG content utilized BT.2020-10. Static title screens containing the sequence name and duration were trimmed from the beginning of sequences to prevent skewing Rate-Distortion results.

\vspace{-3.5mm}
\subsubsection{Encoding Configuration}
\vspace{-1mm}

All experimental encodings were performed on an NVIDIA RTX PRO 6000 Blackwell Max-Q Workstation Edition GPU. The specific system configuration is detailed in \cref{tab:hardware}. We utilized a custom build of FFmpeg based on version \textit{N-122268-g0dfaed77a6} patched with our Pull Request \#21371, which implemented the API calls to introduce support for 4-way SFE and has since been merged into the mainline codebase \cite{ffmpeg_2026}. We automated the testing of various presets and SFE split modes using a PowerShell script (see \textit{Supplementary Material}). The encoding parameters were selected to mirror strict broadcast contribution standards as outlined in recent work and whitepapers \cite{11417632,twitch,google_2019}:

\begin{itemize}
  \item \textbf{Rate Control:} Constant Bitrate (CBR) with a Video Buffer Verifier (VBV) buffer size set to 2× the target bitrate.
  \item \textbf{GOP Structure:} Fixed 2-second Group of Pictures.
  \item \textbf{Frame Structure:} 2 B-frames and 1 Reference frame (\textit{-bf 2 -refs 1}).
  \item \textbf{Format:} HEVC (H.265) Main 10 profile, 4:2:2 chroma subsampling, 10-bit depth.
  \item \textbf{Optimization:} Spatial and Temporal Adaptive Quantization were disabled (\textit{-spatial\_aq 0 -temporal\_aq 0}). Lookahead was disabled (\textit{-rc-lookahead 0}) to minimize latency.
\end{itemize}

We encoded to ten distinct bitrate points for each sequence, as listed in \cref{tab:bitrateRange}. All Rate-Distortion benchmarks were conducted using the \textit{High Quality (HQ)} tuning configuration. As established in recent comparative studies \cite{11417632}, \textit{Low Latency (LL)} tuning on NVENC hardware imposes measurable rate-distortion penalties (approximately -0.2 dB Peak Signal-to-Noise Ratio or PSNR) while offering negligible end-to-end latency reductions for 8K workflows, unless sub-frame latency (\textless16 ms) is strictly required.
\vspace{-3.5mm}
\subsubsection{Quality Metrics}
\vspace{-1.5mm}

\begin{table}[!tbp]
\vspace{-2.5mm}
\setstretch{0.85}
\caption{Target Encoding Bitrates for each resolution}
\centering
\label{tab:bitrateRange}
\resizebox{7.5cm}{!}{\begin{tabular}{@{}lc@{}}
\toprule
Resolution                 & Bitrates (Mbps) \\\midrule
3840×2160 (2160p/4K) & 1, 2, 3, 4, 7, 10, 15, 22, 35, 50 \\
7680×4320 (4320p/8K) & 4, 6, 9, 14, 20, 30, 45, 70, 100, 150 \\
\bottomrule
\end{tabular}}
\vspace{-6.5mm}
\end{table}

\begin{table*}[!tbp]
\setstretch{0.60}
\caption{Encoding Throughput in Frames Per Second (fps) for each combination of Resolution, Codec, Preset, and Tuning.}
\centering
\label{tab:EncodingThroughput}
\resizebox{16cm}{!}{\begin{tabular}{@{}ccccccccccccccc@{}}
\toprule
\multirow{2.5}{*}{Resolution} & \multirow{2.5}{*}{Tuning} & \multirow{2.5}{*}{Preset} & \multicolumn{4}{c}{HEVC (H.265) 8-Bit 4:2:0} & \multicolumn{4}{c}{HEVC (H.265) 10-Bit 4:2:2} & \multicolumn{4}{c}{AV1 8-Bit 4:2:0}\\
\cmidrule(lr){4-7} \cmidrule(lr){8-11} \cmidrule(lr){12-15}
 &&&1 Way & 2 Ways & 3 Ways & 4 Ways & 1 Way & 2 Ways & 3 Ways & 4 Ways & 1 Way & 2 Ways & 3 Ways & 4 Ways \\
\midrule
\multirow{10}{*}{4K UHD} & \multirow{4.5}{*}{High Quality (HQ)} & P1&255&476&634&779&230&431&575&713&237&440&599&734 \\
\cmidrule(lr){3-15}
&&P4&126&244&333&423&126&242&331&423&123&238&329&417 \\
\cmidrule(lr){3-15}
&&P7&39&77&108&143&41&80&113&150&92&180&251&320      \\
\cmidrule(lr){2-15}
&\multirow{4.5}{*}{Low Latency (LL)}&P1&255&475&632&780&230&431&575&714&251&465&633&771 \\
\cmidrule(lr){3-15}
&&P4&126&244&333&424&126&242&331&423&173&327&448&557 \\
\cmidrule(lr){3-15}
&&P7&112&218&299&383&111&217&297&382&97&189&263&335  \\\midrule
\multirow{10}{*}{8K UHD} & \multirow{4.5}{*}{High Quality (HQ)} & P1&66&130&188&247&59&115&166&218&61&119&172&225    \\
\cmidrule(lr){3-15}
&&P4&32&63&92&124&32&62&90&122&31&61&90&120          \\
\cmidrule(lr){3-15}
&&P7&10&20&29&39&10&21&30&41&23&46&67&88             \\
\cmidrule(lr){2-15}
&\multirow{4.5}{*}{Low Latency (LL)}&P1&66&130&188&247&59&115&166&218&64&127&183&238    \\
\cmidrule(lr){3-15}
&&P4&32&63&92&124&32&62&91&122&44&85&126&166         \\
\cmidrule(lr){3-15}
&&P7&29&56&82&111&28&55&81&109&25&48&71&93           \\

\bottomrule
\end{tabular}}
\vspace{-6mm}
\end{table*}

Rate-distortion performance was analyzed using Bjøntegaard Delta Rate (BD-Rate) \cite{barman2024bjontegaarddeltabdtutorial} derived from the ten bitrate points. We computed PSNR, Structural Similarity Index Measure (SSIM), and Video Multimethod Assessment Fusion (VMAF) for all encodes. For PSNR, the standard FFmpeg PSNR filter outputs infinite values for identical (black) frames, which result in biased average scores. To correct for this in 10-bit content, we capped the maximum PSNR at 60.198 dB for each frame (the theoretical limit for 10-bit video) before calculating the sequence average \cite{waveletbeam_2019}.

On the other hand, for VMAF, we utilized the \textit{vmaf\_4k\_v0.\allowbreak6.1neg} model, which is optimized to measure pure compression efficiency \cite{li_swanson_bampis_krasula_aaron_2020}. As the current VMAF model is trained on 8-bit 4:2:0 sources, we utilized the FFmpeg filters to convert both the distorted and reference signals to \textit{yuv420p} prior to calculation. We acknowledge this format conversion is a limitation; therefore, while BD-Rate VMAF results are provided in the \textit{Supplementary Material} for reference, the primary analysis in this paper focuses on BD-Rate PSNR-Y and BD-Rate SSIM-Y. Full results for chroma components (U/V) are also available in the \textit{Supplementary Material}.

For datasets comprising multiple distinct sequences, such as Netflix Chimera, the reported metrics represent the average across all sequences within that dataset. However, for the ITE Ultra-High Definition dataset, the results for the ten 4K sequences and the eleven 8K sequences are reported separately to distinguish the characteristics specific to each resolution. A detailed breakdown of BD-Rate for each sequence is also available in the \textit{Supplementary Material}.

\vspace{-3mm}
\subsection{Power Consumption Analysis}
\vspace{-1mm}
To assess the power characteristics of the Blackwell architecture under sustained encoding loads, we monitored the Board Power Draw and GPU Chip Power Draw during the encoding sessions. Telemetry data was logged using GPU-Z version 2.68.0. For each configuration, the encoding workload was sustained for a duration of one minute at 8K resolution (CBR 200 Mbps, P7) to ensure measurement consistency. Power draw measurements were recorded at 1-second intervals, from which the average board power consumption was calculated. To understand the impact of different codecs and chroma subsampling formats on power draw, three codec and format configurations were evaluated: HEVC (H.265) 8-bit 4:2:0, HEVC (H.265) 10-bit 4:2:2, and AV1 8-bit 4:2:0.
\vspace{-4mm}
\section{Results and Analysis}

\vspace{-1mm}

\subsection{Encoding Throughput}

\vspace{-1mm}

\cref{tab:EncodingThroughput} presents the detailed encoding throughput results. As mentioned earlier, for this analysis, we focus primarily on the \textit{High Quality (HQ)} tuning, as it represents the optimal option for professional broadcasting environments. From the results, it was found that by enabling 4-way SFE, a commercial GPGPU can achieve encoding throughput exceeding the critical threshold of the broadcast industry at 8K 120fps. When encoding to HEVC 10-bit 4:2:2 at 8K resolution using a single encoder (1-Way), the GPU only yields 32 fps at the P4 (Medium) preset, which is insufficient for standard 60fps broadcast, let alone HFR applications. However, enabling 4-Way SFE results in a near-linear performance scaling at 3.81× (see \cref{fig:8KSpeedUp}), boosting throughput to 122 fps. This confirms high-end Blackwell architectures with four NVENC engines effectively support real-time 8K 120fps encoding. Furthermore, 2-way SFE achieves 115 fps at the P1 preset, satisfying 100fps PAL standards. Our recent work indicates smaller GPU dies suffer a smaller internal latency penalty than their larger counterparts, suggesting even better performance, while only single-NVENC 60-class GPUs fall short of these targets \cite{sustainablerealtime8k60hevc}. Furthermore, for scenarios prioritizing speed over chroma information, such as high-speed capture for slow-motion replays, the GPU demonstrates immense headroom. When encoding HEVC 8-bit 4:2:0 at the P1 (Fastest) preset with 4-Way SFE, the GPU yields encoding throughput of 247 fps, enabling real-time 8K 240fps workflows.

\begin{table}[!tbp]
\vspace{-3mm}
\setstretch{0.75}
\caption{BD-Rate (\%) for each SFE configuration relative to No-SFE in HEVC (H.265) 10-Bit 4:2:2}
\centering
\label{tab:BDRate}
\resizebox{8.5cm}{!}{\begin{tabular}{@{}lccccccc@{}}
\toprule
\multirow{3.5}{*}{Dataset} & \multirow{3.5}{*}{Preset} & \multicolumn{2}{c}{2 Ways} & \multicolumn{2}{c}{3 Ways} & \multicolumn{2}{c}{4 Ways}\\
\cmidrule(lr){3-4} \cmidrule(lr){5-6} \cmidrule(lr){7-8}
&&BD-Rate&BD-Rate&BD-Rate&BD-Rate&BD-Rate&BD-Rate\\
&&PSNR-Y&SSIM-Y&PSNR-Y&SSIM-Y&PSNR-Y&SSIM-Y\\
\midrule
\multirow{3}{*}{ITE UHD Series A 4K} 
&P1&0.93&0.82&2.48&2.26&3.13&3.10  \\
&P4&1.97&1.37&3.17&2.35&4.15&3.51  \\
&P7&1.52&1.02&2.61&1.94&4.12&3.78  \\
\midrule
\multirow{3}{*}{ITE UHD Series A 8K} 
&P1&1.60&0.39&2.01&1.26&2.48&1.56  \\
&P4&1.51&0.25&2.17&0.80&3.01&1.46  \\
&P7&1.37&0.05&1.95&0.46&2.21&0.70  \\
\midrule
\multirow{3}{*}{Netflix Sol Levante} 
&P1&0.04&3.56&1.10&4.79&2.18&5.88  \\
&P4&0.18&4.00&1.01&4.93&1.97&5.92  \\
&P7&0.12&3.96&1.09&4.96&1.98&6.19  \\
\midrule
\multirow{3}{*}{Netflix Meridian} 
&P1&-1.03&1.30&0.30&2.89&1.92&3.57 \\
&P4&-0.61&2.80&0.72&4.12&2.15&5.12 \\
&P7&0.01&2.53&1.05&4.23&2.51&4.87  \\
\midrule
\multirow{3}{*}{Netflix Nocturne} 
&P1&2.42&2.42&4.98&4.85&7.21&7.19  \\
&P4&3.82&3.34&6.34&5.67&8.89&8.26  \\
&P7&3.73&3.31&6.13&5.56&8.71&8.02  \\
\midrule
\multirow{3}{*}{Netflix Chimera} 
&P1&2.72&2.39&4.01&3.02&5.14&4.59  \\
&P4&2.81&1.86&4.45&3.68&5.98&5.03  \\
&P7&2.86&2.00&4.44&3.52&6.02&5.66  \\
\midrule
\multirow{3}{*}{\textbf{Average}} 
&P1&1.11&1.81&2.48&3.18&3.68&4.31  \\
&P4&1.61&2.27&2.98&3.59&4.36&4.88  \\
&P7&1.60&2.15&2.88&3.45&4.26&4.87  \\
\bottomrule
\end{tabular}}
\vspace{-6.5mm}
\end{table}

While proprietary implementations prevent verifying the exact slice-boundary mechanisms, the observed near-linear throughput scaling suggests minimal to no cross-slice overlapping is performed. At 4K resolution, the scaling factor is slightly less linear than at 8K (increasing from 126 fps to 423 fps, a 3.35× speedup, see \cref{fig:4KSpeedUp}). This deviation stems from synchronization constraints inherent to the split-frame architecture. Since the completion time of a full frame is bound by its slowest slice, variation in macroblock complexity can cause faster engines to stall while waiting for the busiest engine to finish, an imbalance that is statistically more significant given the shorter frame times at 4K. However, despite this overhead, 4-Way SFE enables the use of the P7 (Slowest) preset to achieve 150 fps in HEVC 10-bit 4:2:2. Hence, for 4K 120fps production, broadcasters can utilize the highest quality settings without compromising real-time stability, a feat not possible with the single-encoder throughput of 41 fps. When the encoding throughput of HEVC 8-bit 4:2:0 is compared to that of HEVC 10-bit 4:2:2, the additional chroma information and bit depth result in a throughput penalty of approximately 12\% when using the P1 preset. However, this penalty was non-uniform; the impact on presets with higher encoding complexity, such as P4 and P7, was negligible. Conversely, when compared to AV1 8-bit 4:2:0, the encoding throughput at the P1 preset was similar to that of HEVC 10-bit 4:2:2. Throughput at the P4 preset was nearly identical across all HEVC and AV1 configurations, while at the P7 preset, AV1 exhibited more than double the throughput of its HEVC counterparts. \looseness=-3

\vspace{-4mm}

\subsection{Rate-Distortion (RD)}
\vspace{-1mm}

\cref{tab:BDRate} details the BD-Rate performance, quantifying the bitrate overhead required to maintain equivalent objective quality when enabling SFE. In terms of BD-Rate PSNR-Y, the results indicate a direct correlation between the number of splits and the loss of coding efficiency. As the frame is being split from 1-Way to 4-Way, the inability to exploit vertical spatial redundancies and Motion Vector (MV) prediction across slice boundaries necessitates a higher bitrate for image reconstruction. For the target P4 (Medium) preset, the average BD-Rate penalty for 4-Way SFE is 4.36\%. Interestingly, the penalty is slightly lower for the P1 (Fastest) preset at 3.68\%. This suggests that the more complex presets are more sensitive to the slicing imposed by SFE than simpler algorithms used in high-speed presets are. The 2-Way split previously evaluated in the literature on Ada Lovelace \cite{11417632} incurs an average BD-Rate penalty ranging from 1.11\% to 1.61\% across the presets, which aligns with recent findings of negligible visual impact. 

As for BD-Rate SSIM-Y, which often correlates more closely with the structural perceptual quality than PSNR, it exhibits a similar trend. The average BD-Rate penalty for preset P4 in a 4-Way configuration is 4.88\%, while the fastest preset P1 yields a slightly lower 4.31\% penalty. This increase indicates that the stitching seams where independent slices meet may introduce minor structural inconsistencies that require additional bits to resolve transparently. Despite this, considering both metrics for a 4-Way configuration at preset P4, we observed average and maximum overheads of roughly 5\% and 9\%, respectively -- a necessary trade-off for the near-quadrupled throughput required to enable real-time 8K 120fps workflows.

Crucially, the RD impact of SFE is highly content-dependent and reveals a divergence between metrics. Animation content, such as \textit{Netflix Sol Levante}, is exceptionally resilient in terms of BD-Rate PSNR-Y with a minimal penalty of 1.97\% (P4, 4-Way), likely due to large flat regions that can be easily exploited intra-slice. However, it suffers a high BD-Rate SSIM-Y penalty of 5.92\%, which suggests that while pixel-level fidelity remains high, the structural interruptions caused by slice boundaries, such as cutting through clean line art or gradients, are penalized more heavily by perceptual metrics in synthetic content than in natural video. In contrast, the penalty is most pronounced in sequences exhibiting a high discrepancy in complexity between the horizontal slices. This occurs when one or two slices contain high-motion or high-texture detail while the remaining slices are largely static, black, or blown out (solid white). Sequences like \textit{Netflix Nocturne} and \textit{Chimera}, which often feature such compositional imbalances, exhibit the highest penalties in both metrics. For example, Nocturne incurs a substantial penalty of 8.89\% in 4-Way mode in terms of BD-Rate PSNR-Y, as its dynamic lighting often creates scenarios where one slice has high complexity while others are nearly static. Conversely, content where complexity is more uniformly distributed across all slices tends to incur a lower penalty. \looseness=-1

\vspace{-4mm}

\subsection{Power Consumption}
\vspace{-1mm}
\begin{table}[!tbp]
\vspace{-2mm}
\setstretch{0.75}
\vspace{-0.5mm}
\caption{Power Consumption (W) by \# of Active NVENC}
\centering
\label{tab:PowerConsumption}
\resizebox{8.5cm}{!}{\begin{tabular}{@{}lccccc@{}}
\toprule
Codec/Format                 & 1 Way & 2 Ways & 3 Ways & 4 Ways & Avg./NVENC \\\midrule
\multicolumn{6}{c}{Board Power Draw (W)}\\
\midrule
HEVC (H.265) 8-Bit 4:2:0&89.4&91.5&92.5&93.4&1.33\\
HEVC (H.265) 10-Bit 4:2:2&91.9&93.4&95.2&99.3&2.47\\
AV1 8-Bit 4:2:0&92.2&93.3&96&100.1&2.63\\
\midrule
\multicolumn{6}{c}{GPU Chip Power Draw (W)}\\
\midrule
HEVC (H.265) 8-Bit 4:2:0&34.4&35.2&35.6&36&0.53\\
HEVC (H.265) 10-Bit 4:2:2&35.3&35.9&36.6&38.2&0.97\\
AV1 8-Bit 4:2:0&35.5&35.9&36.9&38.5&1.00\\
\bottomrule
\end{tabular}}
\vspace{-7mm}
\end{table}
One of the important advantages of the ASIC-based NVENC approach over software or FPGA solutions is energy efficiency. \cref{tab:PowerConsumption} details the power draw during sustained 8K encoding loads. The results reveal a key architectural behavior: activating the NVENC hardware also requires the GPU’s Streaming Multiprocessor (SM) cores to be brought out of their low-power states (P8), resulting in a significant baseline power floor due to leakage power. As seen from the GPU Chip Power Draw, each additional NVENC engine engaged for SFE only slightly increased the power consumption. For the target HEVC 10-bit 4:2:2 encoding, scaling from one to four encoders adds only 2.9 W to the chip's power draw, averaging about 0.97 W per additional active encoder. This efficiency is reflected in the total Board Power Draw, which includes VRAM and other components, increasing by only 7.4 W when moving from a single encoder to a 4-Way SFE configuration. Moreover, it was found that HEVC 8-bit 4:2:0 is the most energy-efficient format, drawing 93.4 W at the board level in 4-Way mode. The move to the professional HEVC 10-bit 4:2:2 encoding increases power consumption to 99.3 W. Finally, the more computationally intensive AV1 codec represents the highest power draw, albeit by a small margin, at 100.1 W. This predictable power scaling confirms that while the baseline power to activate the GPU is substantial, the incremental cost of full parallel utilization is minimal, making 4-Way SFE an efficient operational model.

\vspace{-5mm}

\section{Conclusion and Future Work}
\vspace{-3mm}

This paper presented a comprehensive evaluation of hardware-accelerated 10-bit 4:2:2 video encoding on commercial off-the-shelf (COTS) GPUs, targeting the strict computational demands of real-time V-PCC and next-generation volumetric video streaming. Historically restricted to specialized ASICs, this high-fidelity encoding process is required to prevent geometric deformations, surface tearing, and color bleeding during 3D reconstruction.

Using the NVIDIA RTX PRO 6000 Blackwell Workstation GPU as a testbed, we successfully quantified the throughput, coding efficiency, and power characteristics of a modern GPU hardware video encoder in this workload. Our experimental results confirm that multi-encoder COTS hardware can now natively execute the 10-bit 4:2:2 HEVC encoding process at the speeds necessary to support high-density, dynamic point clouds. By leveraging the multiple encoders, the pipeline achieved an 8K throughput of 122 fps, easily satisfying real-time constraints. While enabling the parallel architectures required to hit these extreme pixel rates introduces a modest BD-Rate penalty of approximately 4.36\% (PSNR-Y) and 4.88\% (SSIM-Y), the preserved structural and color integrity achieved by the 10-bit 4:2:2 encoding process far outweighs these compression overheads for V-PCC applications. Furthermore, our power analysis demonstrated that this professional encoding pipeline operates with remarkable energy efficiency, drawing only 99.3 W under 8K workloads.

As volumetric media scales in density and framerate, the computational intensity of maintaining this robust encoding process will pose challenges for next-generation codecs like VVC. Future work will focus on optimizing 10-bit 4:2:2 encoding pipelines and evaluating V-PCC datasets to determine whether theoretical slice-boundary penalties measurably affect 3D point cloud reconstruction. Confirming that these penalties remain negligible would further establish COTS hardware as a viable baseline for real-time volumetric production. \looseness=-1

\vspace{-1.5mm}

\setstretch{0.8}
\newcommand{\BIBdecl}{\setlength{\itemsep}{0.1 em}} 
\bibliographystyle{IEEEtran}
\bibliography{refs}\looseness=-1

@IEEEtranBSTCTL{IEEEexample:BSTcontrol,
  CTLuse_url = "no",
  CTLuse_forced_etal       = "yes",
  CTLmax_names_forced_etal = "2",
  CTLnames_show_etal       = "2" 
}

@INPROCEEDINGS{11396901,
  author={Arunruangsirilert, Kasidis and Katto, Jiro},
  booktitle={2025 International Conference on Visual Communications and Image Processing (VCIP)}, 
  title={Evaluation of GPU Video Encoder for Low-Latency Real-Time 4K UHD Encoding}, 
  year={2025},
  volume={},
  number={},
  pages={1-5},
  doi={10.1109/VCIP67698.2025.11396901}}

@INPROCEEDINGS{10637525,
  author={Arunruangsirilert, Kasidis and Katto, Jiro},
  booktitle={2024 33rd International Conference on Computer Communications and Networks (ICCCN)}, 
  title={Evaluation of Hardware-based Video Encoders on Modern GPUs for UHD Live-Streaming}, 
  year={2024},
  volume={},
  number={},
  pages={1-9},
  doi={10.1109/ICCCN61486.2024.10637525}}

@ARTICLE{10470357,
  author={Zhang, Junteng and Zhang, Junzhe and Ding, Dandan and Ma, Zhan},
  journal={IEEE Trans. Vis. Comput. Graph.}, 
  title={Learning to Restore Compressed Point Cloud Attribute: A Fully Data-Driven Approach and a Rules-Unrolling-Based Optimization}, 
  year={2025},
  volume={31},
  number={4},
  pages={1985-1998},
  doi={10.1109/TVCG.2024.3375861}}

@misc{sustainablerealtime8k60hevc,
      title={Sustainable Real-Time 8K60 HEVC Encoding for V2X: Repurposing Legacy NVENC Hardware at the Vehicular Edge}, 
      author={Kasidis Arunruangsirilert and Jiro Katto},
      year={2026},
      eprint={2605.16738},
      archivePrefix={arXiv},
      primaryClass={eess.IV},
      url={https://arxiv.org/abs/2605.16738}, 
}

@ARTICLE{10682566,
  author={Wang, Jianqiang and Xue, Ruixiang and Li, Jiaxin and Ding, Dandan and Lin, Yi and Ma, Zhan},
  journal={IEEE Transactions on Pattern Analysis and Machine Intelligence}, 
  title={A Versatile Point Cloud Compressor Using Universal Multiscale Conditional Coding – Part II: Attribute}, 
  year={2025},
  volume={47},
  number={1},
  pages={252-268},
  doi={10.1109/TPAMI.2024.3462945}}

@ARTICLE{10403987,
  author={Ak, Ali and Zerman, Emin and Quach, Maurice and Chetouani, Aladine and Smolic, Aljosa and Valenzise, Giuseppe and Le Callet, Patrick},
  journal={IEEE Transactions on Multimedia}, 
  title={BASICS: Broad Quality Assessment of Static Point Clouds in a Compression Scenario}, 
  year={2024},
  volume={26},
  number={},
  pages={6730-6742},
  doi={10.1109/TMM.2024.3355642}}

@book{nvidia_bw_2024, title={NVIDIA RTX BLACKWELL GPU ARCHITECTURE Built for Neural Rendering ii NVIDIA RTX Blackwell GPU Architecture}, url={https://images.nvidia.com/aem-dam/Solutions/geforce/blackwell/nvidia-rtx-blackwell-gpu-architecture.pdf}, author={NVIDIA}, year={2024}, month={Mar} }

@INPROCEEDINGS{11396839,
  author={Nakajima, Eiko and Lin, Fangzheng and Arunruangsirilert, Kasidis and Katto, Jiro},
  booktitle={2025 International Conference on Visual Communications and Image Processing (VCIP)}, 
  title={Evaluation of 2D Video Interpolation and Extrapolation Methods for Real-Time V-PCC Error Concealment}, 
  year={2025},
  volume={},
  number={},
  pages={1-5},
  doi={10.1109/VCIP67698.2025.11396839}}

@misc{evolutionnvencefficiencylongitudinal,
      title={Evolution of NVENC Efficiency: A Longitudinal Analysis of HQ and UHQ Tuning Efficiency, Latency and Energy Trade-offs}, 
      author={Kasidis Arunruangsirilert and Jiro Katto},
      year={2026},
      eprint={2605.01187},
      archivePrefix={arXiv},
      primaryClass={eess.IV},
      url={https://arxiv.org/abs/2605.01187}, 
}

@INPROCEEDINGS{11417632,
  author={Arunruangsirilert, Kasidis and Katto, Jiro},
  booktitle={2025 Picture Coding Symposium (PCS)}, 
  title={Evaluation of NVENC Split-Frame Encoding (SFE) for UHD Video Transcoding}, 
  year={2025},
  volume={},
  number={},
  pages={1-5},
  doi={10.1109/PCS65673.2025.11417632}}

@INPROCEEDINGS{9043021,
  author={Iwasaki, Shinya and Lei, Xuying and Chida, Kazuhiro and Sugito, Yasuko and Iguchi, Kazuhisa and Kanda, Kikufumi and Miyoshi, Hidenobu and Uehara, Yoshifumi},
  booktitle={2020 IEEE International Conference on Consumer Electronics (ICCE)}, 
  title={The Required Video Bitrate for 8K120-HZ Real-time Temporal Scalable Coding}, 
  year={2020},
  volume={},
  number={},
  pages={1-5},
  doi={10.1109/ICCE46568.2020.9043021}}

@book{nvidia_2024b, title={NVIDIA RTX PRO BLACKWELL GPU ARCHITECTURE Built for Neural Rendering ii NVIDIA RTX B lackwell GP U Architecture}, url={https://www.nvidia.com/content/dam/en-zz/Solutions/design-visualization/quadro-product-literature/NVIDIA-RTX-Blackwell-PRO-GPU-Architecture-v1.0.pdf}, author={NVIDIA}, year={2024}, month={Mar} }

@misc{ITE_2016, title={Ultra-high definition/wide-color-gamut standard test sequences – Series A}, url={https://www.ite.or.jp/content/test-materials/uhdtv_a/}, journal={Ultra-high definition/wide-color-gamut standard test sequences – Series A}, author={The Institute of Image Information and Television Engineers}, year={2016}, month={Jan} }

@misc{netflix, title={NETFLIX OPEN CONTENT}, url={https://opencontent.netflix.com/}, journal={opencontent.netflix.com}, author={Netflix, Inc.} }

@misc{li_swanson_bampis_krasula_aaron_2020, title={Toward a Better Quality Metric for the Video Community}, url={https://netflixtechblog.com/toward-a-better-quality-metric-for-the-video-community-7ed94e752a30}, journal={Medium}, author={Li, Zhi and Swanson, Kyle and Bampis, Christos and Krasula, Lukas and Aaron, Anne}, year={2020}, month={Dec} }

@misc{ffmpeg_2026, title={avcodec/nvenc: Add 4-Way Multi NVENC Split Frame Encoding (SDK 13.0) in HEVC and AV1 for RTX PRO 6000 Blackwell \#21371}, url={https://code.ffmpeg.org/FFmpeg/FFmpeg/pulls/21371}, journal={FFmpeg.org}, author={FFmpeg}, year={2026}, month={Jan} }

@misc{waveletbeam_2019, title={Corner case is giving wrong VMAF and PSNR values!}, url={https://github.com/Netflix/vmaf/issues/371}, journal={GitHub}, author={waveletbeam}, year={2019}, month={Oct} }

@misc{barman2024bjontegaarddeltabdtutorial,
      title={Bj{\o}ntegaard Delta (BD): A Tutorial Overview of the Metric, Evolution, Challenges, and Recommendations}, 
      author={Nabajeet Barman and Maria G. Martini and Yuriy Reznik},
      year={2024},
      eprint={2401.04039},
      archivePrefix={arXiv},
      primaryClass={cs.MM},
      url={https://arxiv.org/abs/2401.04039}, 
}

@article{10.1145/3682062,
author = {Rudolph, Michael and Schneegass, Stefan and Rizk, Amr},
title = {Transcoding V-PCC Point Cloud Streams in Real-time},
year = {2025},
issue_date = {September 2025},
publisher = {Association for Computing Machinery},
address = {New York, NY, USA},
volume = {21},
number = {9},
issn = {1551-6857},
url = {https://doi.org/10.1145/3682062},
doi = {10.1145/3682062},
journal = {ACM Trans. Multimedia Comput. Commun. Appl.},
month = sep,
articleno = {250},
numpages = {22}
}

@misc{twitch, title={Twitch Help Portal}, url={https://help.twitch.tv/s/article/broadcasting-guidelines?language=en_US}, journal={help.twitch.tv}, author={Twitch} }

@misc{google_2019, title={Choose live encoder settings, bitrates, and resolutions - YouTube Help}, url={https://support.google.com/youtube/answer/2853702}, journal={Google.com}, author={Google}, year={2019} }

@misc{nvidia_2024, title={Video Encoding at 8K60 with Split-Frame Encoding and NVIDIA Ada Lovelace Architecture}, url={https://developer.nvidia.com/blog/video-encoding-at-8k60-with-split-frame-encoding-and-nvidia-ada-lovelace-architecture/}, journal={NVIDIA Technical Blog}, author={NVIDIA}, year={2024}, month={Jan} }

@article{10.1145/3690641,
author = {Lin, Ting-Lan and Su, Bing-Wei and Shen, Po-Cheng and Chen, Ding-Yuan and Liang, Chi-Fu and Chen, Yan-Cheng and Wen, Yangming and Shahid, Mohammad},
title = {Upsampling Algorithm for V-PCC-Coded 3D Point Clouds},
year = {2024},
issue_date = {December 2024},
publisher = {Association for Computing Machinery},
address = {New York, NY, USA},
volume = {20},
number = {12},
issn = {1551-6857},
url = {https://doi.org/10.1145/3690641},
doi = {10.1145/3690641},
journal = {ACM Trans. Multimedia Comput. Commun. Appl.},
month = nov,
articleno = {367},
numpages = {23}
}

@ARTICLE{10906555,
  author={Li, Wei and Huang, Leilei and He, Chenlong and Jing, Minge and Hu, Wei and Fan, Yibo},
  journal={IEEE Transactions on Circuits and Systems for Video Technology}, 
  title={An 8K@120fps Advanced Entropy Coding Hardware Design for AVS3}, 
  year={2025},
  volume={35},
  number={8},
  pages={8372-8376},
  doi={10.1109/TCSVT.2025.3546486}}

@INPROCEEDINGS{9013947,
  author={Kobayashi, Daisuke and Nakamura, Ken and Osawa, Tatsuya and Omori, Yuya and Onishi, Takayuki and Iwasaki, Hiroe},
  booktitle={2019 IEEE Global Communications Conference (GLOBECOM)}, 
  title={A Real-Time 4K HEVC Multi-Channel Encoding System with Content-Aware Bitrate Control}, 
  year={2019},
  volume={},
  number={},
  pages={1-6},
  doi={10.1109/GLOBECOM38437.2019.9013947}}

@ARTICLE{8388869,
  author={Onishi, Takayuki and Sano, Takashi and Nishida, Yukikuni and Yokohari, Kazuya and Nakamura, Ken and Nitta, Koyo and Kawashima, Kimiko and Okamoto, Jun and Ono, Naoki and Sagata, Atsushi and Iwasaki, Hiroe and Ikeda, Mitsuo and Shimizu, Atsushi},
  journal={IEEE Transactions on Very Large Scale Integration (VLSI) Systems}, 
  title={A Single-Chip 4K 60-fps 4:2:2 HEVC Video Encoder LSI Employing Efficient Motion Estimation and Mode Decision Framework With Scalability to 8K}, 
  year={2018},
  volume={26},
  number={10},
  pages={1930-1938},
  doi={10.1109/TVLSI.2018.2842179}}

\newpage

\appendix

\onecolumn

\section{Encoding Throughput Benchmark Script}

The batch scripts included below are for benchmarking encoding throughput at 8K UHD resolution. For 4K UHD, use the following settings:
\begin{itemize}
  \item set "RES=3840x2160"
  \item -b:v 50M -maxrate 50M -bufsize 100M
  \item -frames:v 14400
\end{itemize}

\subsection{HEVC 8-Bit 4:2:0}

\begin{lstlisting}[language=batch]
@echo off
setlocal enabledelayedexpansion

:: Output log file
set "LOGFILE=nvenc_throughput_test-8K-8bit.log"
set "RES=7680x4320"

echo NVENC RAM-Buffer Speed Test (No Decoder) Started > "%LOGFILE%"
echo Resolution: %RES% >> "%LOGFILE%"

:: =========================================================
:: TEST SET 1: 8-bit 4:2:0
:: =========================================================
echo. >> "%LOGFILE%"
echo ======================================================= >> "%LOGFILE%"
echo [TEST SET 1] 8-bit 4:2:0 (RAM Buffer -> PCIe -> NVENC) >> "%LOGFILE%"
echo ======================================================= >> "%LOGFILE%"
for %%P in (p1 p4 p7) do (
    for %%T in (ll hq) do (
        for %%S in (15 2 3 4) do (
            echo Testing 8-bit 4:2:0 - Preset: %%P - Tune: %%T - SFE: %%S
            echo [8-BIT] Preset: %%P Tune: %%T SFE: %%S >> "%LOGFILE%"
            
            ffmpeg.exe -hide_banner -y ^
            -init_hw_device cuda=cuda_dev:0 -filter_hw_device cuda_dev ^
            -f lavfi -i color=c=black:s=%RES%:r=60 ^
            -c:v hevc_nvenc -preset %%P -tune %%T ^
            -vf "format=yuv420p,trim=end_frame=1,hwupload=extra_hw_frames=64, loop=loop=-1:size=1:start=0" ^
            -spatial_aq 0 -temporal_aq 0 -rc-lookahead 0 -split_encode_mode %%S ^
            -profile:v main -b:v 200M -maxrate 200M -bufsize 400M -rc cbr ^
            -g 120 -bf 2 -refs 1 -b_ref_mode middle ^
            -frames:v 3600 ^
            -f null NUL 2>> "%LOGFILE%"
            echo --------------------------------------------------- >> "%LOGFILE%"
        )
    )
)

echo.
echo Tests Complete. Results saved in %LOGFILE%
pause
\end{lstlisting}
\subsection{HEVC 10-Bit 4:2:2}
\begin{lstlisting}[language=batch]
@echo off
setlocal enabledelayedexpansion

:: Output log file
set "LOGFILE=nvenc_throughput_test-8K-10bit.log"
set "RES=7680x4320"

echo NVENC RAM-Buffer Speed Test (No Decoder) Started > "%LOGFILE%"
echo Resolution: %RES% >> "%LOGFILE%"

:: =========================================================
:: TEST SET 2: 10-bit 4:2:2
:: (Warning: High PCIe Bandwidth Usage ~13GB/s)
:: =========================================================
echo. >> "%LOGFILE%"
echo ======================================================= >> "%LOGFILE%"
echo [TEST SET 2] 10-bit 4:2:2 (RAM Buffer -> PCIe -> NVENC) >> "%LOGFILE%"
echo ======================================================= >> "%LOGFILE%"
for %%P in (p1 p4 p7) do (
    for %%T in (ll hq) do (
        for %%S in (15 2 3 4) do (
            echo Testing 10-bit 4:2:2 - Preset: %%P - Tune: %%T - SFE: %%S
            echo [10-BIT] Preset: %%P Tune: %%T SFE: %%S >> "%LOGFILE%"
            
            ffmpeg.exe -hide_banner -y ^
            -init_hw_device cuda=cuda_dev:0 -filter_hw_device cuda_dev ^
            -f lavfi -i "color=c=black:s=%RES%:r=60" ^
            -c:v hevc_nvenc -preset %%P -tune %%T ^
            -vf "format=p216le,trim=end_frame=1,hwupload=extra_hw_frames=64, loop=loop=-1:size=1:start=0" ^
            -spatial_aq 0 -temporal_aq 0 -rc-lookahead 0 -split_encode_mode %%S ^
            -b:v 200M -maxrate 200M -bufsize 400M -rc cbr ^
            -profile:v main10 -g 120 -bf 2 -refs 1 -b_ref_mode middle ^
            -color_primaries bt2020 -color_trc smpte2084 -colorspace bt2020nc ^
            -frames:v 3600 ^
            -f null NUL 2>> "%LOGFILE%"
            echo --------------------------------------------------- >> "%LOGFILE%"
        )
    )
)

echo.
echo Tests Complete. Results saved in %LOGFILE%
pause
\end{lstlisting}

\subsection{AV1 8-Bit 4:2:0}
\begin{lstlisting}[language=batch]
@echo off
setlocal enabledelayedexpansion

:: Output log file
set "LOGFILE=nvenc_throughput_test-8K-AV1.log"
set "RES=7680x4320"

echo NVENC RAM-Buffer Speed Test (No Decoder) Started > "%LOGFILE%"
echo Resolution: %RES% >> "%LOGFILE%"

:: =========================================================
:: TEST SET 3: 8-bit 4:2:0
:: =========================================================
echo. >> "%LOGFILE%"
echo ======================================================= >> "%LOGFILE%"
echo [TEST SET 3] 8-bit 4:2:0 (RAM Buffer -> PCIe -> NVENC) >> "%LOGFILE%"
echo ======================================================= >> "%LOGFILE%"
for %%P in (p1 p4 p7) do (
    for %%T in (ll hq) do (
        for %%S in (15 2 3 4) do (
            echo Testing 8-bit 4:2:0 - Preset: %%P - Tune: %%T - SFE: %%S
            echo [8-BIT] Preset: %%P Tune: %%T SFE: %%S >> "%LOGFILE%"
            
            ffmpeg.exe -hide_banner -y ^
            -init_hw_device cuda=cuda_dev:0 -filter_hw_device cuda_dev ^
            -f lavfi -i color=c=black:s=%RES%:r=60 ^
            -c:v av1_nvenc -preset %%P -tune %%T ^
            -vf "format=yuv420p,trim=end_frame=1,hwupload=extra_hw_frames=64, loop=loop=-1:size=1:start=0" ^
            -spatial_aq 0 -temporal_aq 0 -rc-lookahead 0 -split_encode_mode %%S ^
            -b:v 200M -maxrate 200M -bufsize 400M -rc cbr ^
            -g 120 -bf 2 -refs 1 -b_ref_mode middle ^
            -frames:v 3600 ^
            -f null NUL 2>> "%LOGFILE%"
            echo --------------------------------------------------- >> "%LOGFILE%"
        )
    )
)


echo.
echo Tests Complete. Results saved in %LOGFILE%
pause
\end{lstlisting}

\section{Rate-Distortion Benchmark Script}

The PowerShell script below is for encoding the \textit{Netflix Chimera} dataset. For other datasets, replace the values of \textit{\$inputData} and \textit{\$bitrates} with the corresponding file names, frame rates, and desired bitrates.

\begin{lstlisting}[language=PowerShell]
# --- CONFIGURATION ---

# 1. Concurrency Control
# Monitor VRAM. If "Out of Memory" occurs, lower this number.
$MaxConcurrentJobs = 4

# 2. Source Files & Framerates
$inputData = @(
    @{ Path="1_BAR_SCENE.mp4";  Fps=60000/1001 },
    @{ Path="2_DINNER_SCENE.mp4";  Fps=60000/1001 },
    @{ Path="3_DANCER.mp4";  Fps=60000/1001 },
    @{ Path="4_DANCERS_COUPLE.mp4";  Fps=60000/1001 },
    @{ Path="5_DANCERS_MONTAGE_MIXED.mp4";  Fps=60000/1001 },
    @{ Path="6_ROLLERCOASTER_SEQUENCE.mp4";  Fps=60000/1001 },
    @{ Path="7_ROLLERCOASTER_POV.mp4";  Fps=60000/1001 },
    @{ Path="8_ROLLERCOASTER_PASSENGER.mp4";  Fps=60000/1001 },
    @{ Path="9_TWIRL_RIDE_BOARDWALK.mp4";  Fps=60000/1001 },
    @{ Path="10_NETFLIX_CARD_TWIRL.mp4";  Fps=60000/1001 },
    @{ Path="11_SEASIDE_AND_PIER.mp4";  Fps=60000/1001 },
    @{ Path="12_WIND_AND_NATURE.mp4";  Fps=60000/1001 },
    @{ Path="13_MOUNTAIN_VIEW_W_TILT.mp4";  Fps=60000/1001 },
    @{ Path="14_MOUNTAIN_VIEW_PAN.mp4";  Fps=60000/1001 },
    @{ Path="15_WALK_LIKE_A_MAN.mp4";  Fps=60000/1001 },
    @{ Path="16_TODDLER_AND_FOUNTAIN.mp4";  Fps=60000/1001 },
    @{ Path="17_DRIVING_POV.mp4";  Fps=60000/1001 },
    @{ Path="18_PLANET_MOBILE.mp4";  Fps=60000/1001 },
    @{ Path="19_DOG_PANTS.mp4";  Fps=60000/1001 },
    @{ Path="20_DOG_BARKS.mp4";  Fps=60000/1001 },
    @{ Path="21_RC_AERIAL.mp4";  Fps=60000/1001 },
    @{ Path="22_BASKEYBALL_FREE_THROW.mp4";  Fps=60000/1001 },
    @{ Path="23_BASKEYBALL_GAME.mp4";  Fps=60000/1001 }
)

# 3. Bitrate Ladder
$bitrates = 1, 2, 3, 4, 7, 10, 15, 22, 35, 50

# 4. Matrix Settings
$presets    = "p1", "p4", "p7"
$tunings    = "ull", "ll", "hq"
$splitModes = @(
    @{ Name="1way"; Value=15 },
    @{ Name="2way"; Value=2 },
    @{ Name="3way"; Value=3 },
    @{ Name="4way"; Value=4 }
)

$baseOutputFolder = ""

# --- STEP 1: BUILD THE JOB QUEUE ---

$JobQueue = @()
New-Item -ItemType Directory -Force -Path $baseOutputFolder | Out-Null

Write-Host "Building Job Queue... This may take a moment." -ForegroundColor Cyan

foreach ($item in $inputData) {
    $file = $item.Path
    # Handle FPS calculation
    if ($item.Fps -is [string] -and $item.Fps -match "/") {
        $parts = $item.Fps -split "/"
        $fpsVal = [double]$parts[0] / [double]$parts[1]
    } else {
        $fpsVal = [double]$item.Fps
    }
    
    $gop = [math]::Round($fpsVal * 2)

    if (-not (Test-Path $file)) { continue }
    
    $fileNameWithoutExt = [System.IO.Path]::GetFileNameWithoutExtension($file)
    $sceneFolder = "$baseOutputFolder\$fileNameWithoutExt"
    if (-not (Test-Path $sceneFolder)) { New-Item -ItemType Directory -Force -Path $sceneFolder | Out-Null }

    foreach ($b in $bitrates) {
        $bufSize = $b * 2
        foreach ($preset in $presets) {
            foreach ($tune in $tunings) {
                foreach ($split in $splitModes) {
                    
                    $outputName = "$sceneFolder\${fileNameWithoutExt}_${b}Mbps_${preset}_${tune}_ $($split.Name).mp4"
                    
                    if (-not (Test-Path $outputName)) {
                        $JobQueue += [PSCustomObject]@{
                            InputFile  = $file
                            OutputFile = $outputName
                            Args       = @(
                                "-y",
                                "-hwaccel", "cuda", 
                                "-hwaccel_output_format", "cuda",
                                "-i", "`"$file`"",
                                "-vcodec", "hevc_nvenc",
                                "-preset", "$preset",
                                "-tune", "$tune",
                                "-spatial_aq", "0",
                                "-temporal_aq", "0",
                                "-rc-lookahead", "0",
                                "-split_encode_mode", "$($split.Value)",
                                "-b:v", "${b}M",
                                "-maxrate", "${b}M",
                                "-bufsize", "${bufSize}M",
                                "-rc", "cbr",
                                "-profile:v", "main10",
                                "-g", "$gop",
                                "-surfaces", "64",
                                "-bf", "2",
                                "-refs", "1",
                                "-b_ref_mode", "middle",
                                "-color_primaries", "bt2020",
                                "-color_trc", "smpte2084",
                                "-colorspace", "bt2020nc",
                                "-an",
                                "`"$outputName`""
                            )
                            Description = "$fileNameWithoutExt | ${b}M | $preset | $tune | $($split.Name)"
                        }
                    }
                }
            }
        }
    }
}

$TotalJobs = $JobQueue.Count
Write-Host "Total Jobs Queued: $TotalJobs" -ForegroundColor Green
Start-Sleep -Seconds 2

# --- STEP 2: EXECUTE PARALLEL JOBS ---

# Initialize as an explicit array
$ActiveProcesses = @() 
$CompletedCount  = 0
$NextJobIndex    = 0

while ($CompletedCount -lt $TotalJobs) {
    
    # --- FIX IS HERE ---
    # We wrap the result in @(...) to force it to remain an array 
    # even if filtering leaves 0 or 1 items.
    $ActiveProcesses = @($ActiveProcesses | Where-Object { 
        if ($_.HasExited) {
            $global:CompletedCount++
            return $false # Remove from list
        }
        return $true # Keep in list
    })

    # Launch new jobs if slots are available
    while ($ActiveProcesses.Count -lt $MaxConcurrentJobs -and $NextJobIndex -lt $TotalJobs) {
        $job = $JobQueue[$NextJobIndex]
        
        $p = Start-Process -FilePath "ffmpeg" -ArgumentList $job.Args -NoNewWindow -PassThru
        
        # Now this addition works because $ActiveProcesses is guaranteed to be an array
        $ActiveProcesses += $p
        $NextJobIndex++
    }

    $percent = [math]::Round(($CompletedCount / $TotalJobs) * 100, 1)
    $running = $ActiveProcesses.Count
    
    Write-Progress -Activity "Blackwell Encoding Matrix" `
                   -Status "$percent% Complete ($CompletedCount / $TotalJobs)" `
                   -CurrentOperation "Active Encoders: $running | Queue Index: $NextJobIndex" `
                   -PercentComplete $percent

    Start-Sleep -Milliseconds 250
}

Write-Host "`nAll $TotalJobs encodes completed successfully." -ForegroundColor Green
\end{lstlisting}
\newpage
\section{Full Rate-Distortion Results}

\subsection{Peak Signal-to-Noise Ratio (PSNR)}

\begin{table}[!h]
\vspace{-3mm}
\setstretch{0.65}
\caption{BD-Rate PSNR (\%) for each SFE configuration relative to No-SFE in H.265/HEVC 10-Bit 4:2:2 (ITE UHD Series A)}
\centering
\vspace{2mm}
\resizebox{14.5cm}{!}{\begin{tabular}{@{}lcccccccccc@{}}
\toprule
\multirow{3.5}{*}{Sequence Name} & \multirow{3.5}{*}{Preset} & \multicolumn{3}{c}{2 Ways} & \multicolumn{3}{c}{3 Ways} & \multicolumn{3}{c}{4 Ways}\\
\cmidrule(lr){3-5} \cmidrule(lr){6-8} \cmidrule(lr){9-11}
&&BD-Rate&BD-Rate&BD-Rate&BD-Rate&BD-Rate&BD-Rate&BD-Rate&BD-Rate&BD-Rate\\
&&PSNR-Y&PSNR-U&PSNR-V&PSNR-Y&PSNR-U&PSNR-V&PSNR-Y&PSNR-U&PSNR-V\\
\midrule
\multicolumn{11}{c}{\textbf{ITE UHD Series A 4K}}\\
\midrule
\multirow{3}{*}{a03\_TrainsC\_4K} 
&P1&2.42&2.09&0.75&5.18&3.98&2.69&7.81&7.30&5.76              \\
&P4&4.54&2.77&2.02&6.80&5.17&5.15&8.14&5.09&6.55                            \\
&P7&3.52&2.73&3.03&5.69&4.32&4.29&6.71&5.90&5.48                            \\
\midrule
\multirow{3}{*}{a04\_Expressway\_4K} 
&P1&2.65&4.13&2.05&3.88&6.24&1.52&6.46&9.67&5.10           \\
&P4&3.81&4.73&2.39&6.47&8.87&4.60&8.15&11.62&6.47                           \\
&P7&3.16&3.09&1.61&4.82&6.62&2.11&7.07&9.75&3.51                            \\
\midrule
\multirow{3}{*}{a05\_SteelPlant\_4K} 
&P1&0.22&0.32&-1.72&0.47&0.19&-1.41&0.69&0.88&-0.58        \\
&P4&0.22&0.69&-2.11&0.51&1.48&-1.73&0.74&2.52&-0.73                         \\
&P7&0.21&0.16&-1.65&0.48&1.41&-1.45&0.73&1.63&-0.30                         \\
\midrule
\multirow{3}{*}{a06\_Festival\_4K} 
&P1&0.82&0.99&0.72&1.96&2.48&1.99&2.93&3.70&3.18             \\
&P4&1.10&1.20&0.75&2.13&2.47&1.91&3.10&3.67&2.85                            \\
&P7&1.00&1.05&0.67&2.06&2.34&1.76&3.03&3.67&2.85                            \\
\midrule
\multirow{3}{*}{a07\_River\_4K} 
&P1&0.42&-49.16&37.38&0.66&-36.58&51.91&1.48&-48.50&56.52       \\
&P4&0.44&-48.77&30.24&0.65&-41.27&48.00&1.51&-47.89&53.28                   \\
&P7&0.39&-48.03&20.17&0.62&-36.92&43.22&1.44&-38.85&45.94                   \\
\midrule
\multirow{3}{*}{a08\_JapaneseMaple\_4K} 
&P1&0.64&1.40&0.69&2.29&2.56&1.98&2.46&2.96&1.57        \\
&P4&0.95&1.02&1.47&1.26&1.52&1.33&2.15&2.41&2.26                            \\
&P7&-0.42&-0.65&-0.64&1.85&1.87&1.53&3.36&3.61&2.67                         \\
\midrule
\multirow{3}{*}{a09\_Maiko\_4K} 
&P1&1.29&-0.04&-0.22&2.04&1.50&0.33&2.93&2.63&1.40              \\
&P4&3.52&0.91&0.84&3.07&1.64&-0.22&3.88&2.74&0.39                           \\
&P7&3.18&0.06&-0.38&4.05&2.41&1.01&5.38&4.55&1.79                           \\
\midrule
\multirow{3}{*}{a10\_Umbrella\_4K} 
&P1&1.12&3.83&1.50&2.42&5.75&2.72&1.20&5.93&3.32             \\
&P4&1.62&1.97&1.24&0.91&1.94&1.64&3.40&6.53&4.69                            \\
&P7&0.35&2.29&0.97&1.11&3.57&1.96&3.02&7.32&3.63                            \\
\midrule
\multirow{3}{*}{a11\_LayeredKimono\_4K} 
&P1&-3.21&-0.26&-4.75&1.33&2.18&-0.47&-1.33&3.27&-3.13  \\
&P4&-0.74&-1.28&-2.20&3.31&1.14&1.34&1.04&1.14&0.09                         \\
&P7&1.32&-0.84&1.11&0.03&-1.19&-1.45&2.82&2.48&2.43                         \\
\midrule
\multirow{3}{*}{a12\_Railcar\_4K} 
&P1&2.93&7.22&3.11&4.61&8.40&3.27&6.67&13.81&3.62             \\
&P4&4.28&6.38&3.94&6.57&8.78&4.97&9.40&14.62&8.60                           \\
&P7&2.53&5.08&2.88&5.37&7.94&4.90&7.68&12.16&5.90                           \\
\midrule
\multirow{3}{*}{\textbf{Average}} 
&P1&0.93&-2.95&3.95&2.48&-0.33&6.45&3.13&0.17&7.68  \\
&P4&1.97&-3.04&3.86&3.17&-0.83&6.70&4.15&0.25&8.45  \\
&P7&1.52&-3.51&2.78&2.61&-0.76&5.79&4.12&1.22&7.39  \\
\midrule
\multicolumn{11}{c}{\textbf{ITE UHD Series A 8K}}\\
\midrule
\multirow{3}{*}{a01\_TrainsA\_8K} 
&P1&6.54&4.05&3.94&7.64&4.95&4.17&7.92&5.10&4.44          \\
&P4&6.30&3.63&2.90&7.35&4.12&3.03&10.11&5.96&5.80         \\
&P7&6.60&4.16&3.54&7.26&3.53&2.58&9.46&5.46&5.46          \\
\midrule
\multirow{3}{*}{a02\_TrainsB\_8K} 
&P1&-0.48&0.09&-1.33&0.46&1.35&-0.75&1.08&0.75&-0.18      \\
&P4&0.61&-0.35&-0.86&0.88&0.15&-0.84&1.43&-0.21&-0.34     \\
&P7&0.78&-0.24&-0.96&1.03&0.17&-1.36&1.59&0.00&-0.87      \\
\midrule
\multirow{3}{*}{a04\_Expressway\_8K} 
&P1&4.55&1.26&2.74&4.74&0.73&2.26&4.87&0.35&2.31          \\
&P4&2.87&-1.35&-0.60&5.05&-1.00&1.83&5.99&-0.25&2.06      \\
&P7&3.68&-0.32&1.16&4.47&-1.13&1.43&5.32&-0.54&1.34       \\
\midrule
\multirow{3}{*}{a05\_SteelPlant\_8K} 
&P1&0.54&0.53&-1.01&0.53&0.77&-0.85&0.83&1.57&-0.29       \\
&P4&0.54&0.39&-0.80&0.70&0.84&-0.47&0.83&1.56&-0.38       \\
&P7&0.51&0.47&-0.62&0.67&0.99&-0.28&0.82&1.73&-0.27       \\
\midrule
\multirow{3}{*}{a06\_Festival\_8K} 
&P1&0.44&0.47&0.27&1.03&1.00&0.60&1.57&1.59&1.05          \\
&P4&0.43&0.55&0.22&1.00&1.09&0.59&1.61&1.75&1.10          \\
&P7&0.49&0.65&0.30&1.13&1.24&0.70&1.69&1.87&1.13          \\
\midrule
\multirow{3}{*}{a07\_River\_8K} 
&P1&0.18&-11.83&9.65&0.47&-9.52&2.31&0.49&-14.02&11.52    \\
&P4&0.19&-17.14&7.55&0.48&-7.99&0.86&0.51&-12.10&10.03    \\
&P7&0.20&-13.84&8.45&0.51&-12.78&-2.28&0.46&-8.24&9.27    \\
\midrule
\multirow{3}{*}{a08\_JapaneseMaple\_8K} 
&P1&0.43&1.40&0.71&0.63&1.36&0.79&1.49&2.11&2.16          \\
&P4&1.29&1.77&0.96&1.13&1.78&0.24&1.11&1.88&-0.55         \\
&P7&0.42&0.23&0.44&0.84&0.76&0.09&0.30&0.50&-1.64         \\
\midrule
\multirow{3}{*}{a09\_Maiko\_8K} 
&P1&6.23&0.01&0.27&8.38&0.03&1.77&8.65&1.23&1.27          \\
&P4&4.29&-1.57&-2.69&5.03&-1.38&-2.14&6.29&0.13&-2.39     \\
&P7&4.67&-0.01&-1.41&5.37&-0.80&-1.87&5.15&-0.37&-1.89    \\
\midrule
\multirow{3}{*}{a10\_Umbrella\_8K} 
&P1&-0.23&1.77&1.47&0.02&2.86&1.26&-0.28&2.82&2.44        \\
&P4&0.13&2.29&1.82&0.56&2.86&1.66&2.17&4.87&3.55          \\
&P7&1.16&2.54&2.51&0.87&2.55&1.58&1.61&3.88&2.56          \\
\midrule
\multirow{3}{*}{a11\_LayeredKimono\_8K} 
&P1&-0.63&-0.51&-1.43&-1.83&1.98&-3.18&0.56&5.00&-1.12    \\
&P4&-0.39&-0.97&-1.35&1.25&0.64&0.46&2.79&0.08&0.08       \\
&P7&-3.67&-4.31&-5.00&-1.33&-4.72&-2.14&-2.75&-3.69&-4.53 \\
\midrule
\multirow{3}{*}{a12\_Railcar\_8K} 
&P1&0.07&-0.97&-4.47&0.09&-1.06&-4.38&0.11&-1.20&-4.50    \\
&P4&0.38&-1.10&-4.78&0.39&-1.80&-6.16&0.30&-1.81&-5.83    \\
&P7&0.27&-1.44&-5.46&0.60&-1.18&-5.42&0.64&-1.23&-5.65    \\
\midrule
\multirow{3}{*}{\textbf{Average}} 
&P1&1.60&-0.34&0.98&2.01&0.40&0.36&2.48&0.48&1.74         \\
&P4&1.51&-1.26&0.22&2.17&-0.06&-0.09&3.01&0.17&1.19       \\
&P7&1.37&-1.10&0.27&1.95&-1.03&-0.63&2.21&-0.06&0.45      \\

\bottomrule
\end{tabular}}
\vspace{-6.5mm}
\end{table}

\begin{table}[!h]
\vspace{-3mm}
\setstretch{0.65}
\caption{BD-Rate PSNR (\%) for each SFE configuration relative to No-SFE in H.265/HEVC 10-Bit 4:2:2 (Netflix Chimera)}
\centering
\vspace{2mm}
\resizebox{14.5cm}{!}{\begin{tabular}{@{}lcccccccccc@{}}
\toprule
\multirow{3.5}{*}{Sequence Name} & \multirow{3.5}{*}{Preset} & \multicolumn{3}{c}{2 Ways} & \multicolumn{3}{c}{3 Ways} & \multicolumn{3}{c}{4 Ways}\\
\cmidrule(lr){3-5} \cmidrule(lr){6-8} \cmidrule(lr){9-11}
&&BD-Rate&BD-Rate&BD-Rate&BD-Rate&BD-Rate&BD-Rate&BD-Rate&BD-Rate&BD-Rate\\
&&PSNR-Y&PSNR-U&PSNR-V&PSNR-Y&PSNR-U&PSNR-V&PSNR-Y&PSNR-U&PSNR-V\\
\midrule
\multirow{3}{*}{1\_BAR\_SCENE} 
&P1&1.48&0.41&0.57&2.62&2.79&2.91&3.15&2.59&2.10           \\
&P4&3.55&1.56&1.50&4.10&2.14&2.36&5.47&4.13&4.52           \\
&P7&2.24&0.40&0.85&2.41&1.05&1.88&4.08&3.23&3.73           \\
\midrule
\multirow{3}{*}{2\_DINNER\_SCENE} 
&P1&4.96&7.33&-2.03&5.08&12.78&-1.77&5.91&13.74&2.46       \\
&P4&7.82&11.49&4.06&9.00&8.90&2.96&12.79&17.01&6.39        \\
&P7&7.80&8.56&5.39&9.59&12.01&3.79&10.43&13.05&14.69       \\
\midrule
\multirow{3}{*}{3\_DANCER} 
&P1&3.54&1.57&5.40&3.21&0.58&16.33&4.33&2.12&23.40         \\
&P4&0.19&7.96&27.58&-1.58&15.17&36.61&-1.54&18.65&40.77    \\
&P7&-2.76&10.31&25.54&-1.69&16.40&33.39&-1.70&15.91&39.54  \\
\midrule
\multirow{3}{*}{4\_DANCERS\_COUPLE} 
&P1&5.99&10.42&9.00&-2.90&0.86&11.81&-1.33&5.75&14.57      \\
&P4&2.00&5.10&7.03&3.65&10.26&17.53&4.93&13.20&26.54       \\
&P7&2.26&5.39&7.91&3.61&-22.58&19.80&4.69&-18.34&24.38     \\
\midrule
\multirow{3}{*}{5\_DANCERS\_MONTAGE\_MIXED} 
&P1&-6.44&6.56&27.84&-6.40&6.23&28.80&-5.22&10.05&7.38     \\
&P4&4.33&4.03&-13.03&7.51&5.84&8.15&7.51&6.45&9.39         \\
&P7&3.65&-1.21&8.18&6.19&0.23&-8.50&7.64&0.84&-6.64        \\
\midrule
\multirow{3}{*}{6\_ROLLERCOASTER\_SEQUENCE} 
&P1&3.64&3.60&3.59&6.06&6.39&6.73&8.04&8.71&9.00           \\
&P4&4.14&3.29&3.43&6.45&5.97&6.24&7.98&7.82&8.23           \\
&P7&3.64&3.11&3.14&6.15&5.68&6.33&8.00&8.18&8.57           \\
\midrule
\multirow{3}{*}{7\_ROLLERCOASTER\_POV} 
&P1&4.87&5.07&4.87&7.58&8.18&7.78&9.31&10.24&10.12         \\
&P4&5.00&4.87&4.89&7.82&8.10&8.28&9.34&10.04&10.17         \\
&P7&4.53&4.48&4.39&7.32&7.51&7.69&9.10&9.87&9.84           \\
\midrule
\multirow{3}{*}{8\_ROLLERCOASTER\_PASSENGER} 
&P1&3.29&2.74&2.91&3.97&3.45&3.33&6.16&5.74&5.62           \\
&P4&3.52&2.14&2.26&5.55&4.54&4.75&7.41&6.59&6.91           \\
&P7&3.64&2.65&2.72&5.31&4.37&4.51&7.41&6.85&7.18           \\
\midrule
\multirow{3}{*}{9\_TWIRL\_RIDE\_BOARDWALK} 
&P1&2.33&2.82&2.82&4.30&5.21&5.13&5.97&7.23&7.29           \\
&P4&2.40&2.72&2.91&4.40&5.15&5.03&6.18&7.25&7.30           \\
&P7&2.33&2.72&2.82&4.26&5.02&4.97&5.92&6.99&7.20           \\
\midrule
\multirow{3}{*}{10\_NETFLIX\_CARD\_TWIRL} 
&P1&3.24&3.38&3.23&4.50&5.19&4.96&5.89&7.28&7.10           \\
&P4&2.78&2.91&2.77&3.72&4.26&3.93&5.45&6.56&6.61           \\
&P7&3.40&3.39&3.35&4.72&5.39&5.04&6.16&7.30&7.43           \\
\midrule
\multirow{3}{*}{11\_SEASIDE\_AND\_PIER} 
&P1&2.13&2.30&-0.76&3.17&3.95&0.20&4.67&6.05&1.54          \\
&P4&2.24&0.98&-1.82&3.52&2.99&-0.30&4.54&5.10&1.49         \\
&P7&2.39&2.16&-0.09&3.75&3.59&1.07&4.69&6.02&2.45          \\
\midrule
\multirow{3}{*}{12\_WIND\_AND\_NATURE} 
&P1&0.84&0.28&-0.25&8.44&4.05&1.50&1.82&10.01&11.24        \\
&P4&0.87&-0.30&-0.62&1.01&0.49&-0.91&1.86&1.60&0.42        \\
&P7&0.39&-0.90&-1.27&0.54&0.23&10.85&1.18&0.98&-0.59       \\
\midrule
\multirow{3}{*}{13\_MOUNTAIN\_VIEW\_W\_TILT} 
&P1&1.85&1.98&2.51&4.22&6.01&5.86&8.93&10.88&8.47          \\
&P4&1.65&0.94&2.01&4.45&5.03&5.22&8.03&8.77&8.86           \\
&P7&5.18&3.82&2.08&7.40&6.57&5.59&9.89&9.80&8.62           \\
\midrule
\multirow{3}{*}{14\_MOUNTAIN\_VIEW\_PAN} 
&P1&3.85&3.00&1.67&6.29&4.03&3.81&7.49&8.26&54.36          \\
&P4&3.82&2.72&3.23&5.46&4.29&5.64&7.70&6.41&8.33           \\
&P7&3.35&1.35&1.33&4.64&1.72&2.61&5.65&4.45&6.37           \\
\midrule
\multirow{3}{*}{15\_WALK\_LIKE\_A\_MAN} 
&P1&2.25&1.17&1.17&4.26&4.08&4.47&4.95&5.46&6.00           \\
&P4&2.08&-0.69&0.56&3.62&1.95&4.63&4.21&2.72&5.37          \\
&P7&4.47&-0.44&1.68&3.93&1.29&3.41&7.23&5.19&8.07          \\
\midrule
\multirow{3}{*}{16\_TODDLER\_AND\_FOUNTAIN} 
&P1&0.73&0.44&0.47&1.21&1.53&1.24&1.80&3.84&3.14           \\
&P4&0.94&-0.09&0.81&1.75&0.45&2.66&2.35&3.08&5.14          \\
&P7&0.93&-0.37&0.34&1.59&0.21&2.12&2.32&2.88&4.64          \\
\midrule
\multirow{3}{*}{17\_DRIVING\_POV} 
&P1&2.45&2.43&1.65&4.13&3.96&3.02&5.72&6.16&5.32           \\
&P4&2.41&2.46&1.73&4.08&4.02&3.53&5.85&6.50&5.71           \\
&P7&2.48&2.95&1.89&3.95&4.30&3.51&5.68&6.87&5.88           \\
\midrule
\multirow{3}{*}{18\_PLANET\_MOBILE} 
&P1&1.39&3.44&-2.76&2.43&4.08&2.07&3.37&5.45&7.28          \\
&P4&-3.00&1.44&-5.94&-2.30&3.85&0.91&-2.04&5.54&7.60       \\
&P7&-3.08&0.85&-2.14&-3.12&3.03&-0.29&-1.62&5.47&10.27     \\
\midrule
\multirow{3}{*}{19\_DOG\_PANTS} 
&P1&4.34&2.38&1.66&6.58&4.26&3.64&7.72&6.17&4.88           \\
&P4&3.41&1.33&1.24&5.97&3.97&3.54&7.11&5.22&4.24           \\
&P7&5.75&3.26&3.15&8.37&5.81&5.15&9.24&6.82&5.67           \\
\midrule
\multirow{3}{*}{20\_DOG\_BARKS} 
&P1&2.98&4.45&4.19&3.96&6.83&6.49&6.03&7.88&7.91           \\
&P4&4.09&4.11&3.47&7.00&7.66&6.32&8.22&9.58&8.22           \\
&P7&1.64&1.08&0.74&3.60&3.29&2.40&6.00&7.35&6.70           \\
\midrule
\multirow{3}{*}{21\_RC\_AERIAL} 
&P1&1.73&2.46&4.06&3.40&4.34&6.51&5.12&6.51&9.36           \\
&P4&-1.24&-1.69&-2.90&0.60&0.23&-0.43&1.93&1.91&2.33       \\
&P7&1.93&1.70&2.30&3.85&4.34&5.41&5.14&6.11&8.50           \\
\midrule
\multirow{3}{*}{22\_BASKEYBALL\_FREE\_THROW} 
&P1&6.28&5.22&5.33&7.99&10.03&8.62&8.10&11.48&11.53        \\
&P4&5.94&6.90&6.82&7.63&8.99&8.44&10.91&13.61&13.38        \\
&P7&3.59&2.82&4.31&6.92&7.10&7.24&10.10&11.55&12.26        \\
\midrule
\multirow{3}{*}{23\_BASKEYBALL\_GAME} 
&P1&4.90&4.88&5.01&8.15&8.79&9.00&10.24&11.73&11.80        \\
&P4&5.60&5.68&5.24&8.97&9.75&9.35&11.34&12.96&12.51        \\
&P7&5.98&6.11&5.91&8.90&9.76&9.57&11.15&12.97&12.66        \\
\midrule
\multirow{3}{*}{\textbf{Average}} 
&P1&2.72&3.41&3.57&4.01&5.11&6.19&5.14&7.54&10.08          \\
&P4&2.81&3.04&2.49&4.45&5.39&6.28&5.98&7.86&9.15           \\
&P7&2.86&2.79&3.67&4.44&3.75&5.98&6.02&6.10&9.02           \\
\bottomrule
\end{tabular}}
\vspace{-6.5mm}
\end{table}

\begin{table}[!h]
\vspace{-3mm}
\setstretch{0.65}
\caption{BD-Rate PSNR (\%) for each SFE configuration relative to No-SFE in H.265/HEVC 10-Bit 4:2:2 (Other)}
\centering
\vspace{2mm}
\resizebox{14.5cm}{!}{\begin{tabular}{@{}lcccccccccc@{}}
\toprule
\multirow{3.5}{*}{Sequence Name} & \multirow{3.5}{*}{Preset} & \multicolumn{3}{c}{2 Ways} & \multicolumn{3}{c}{3 Ways} & \multicolumn{3}{c}{4 Ways}\\
\cmidrule(lr){3-5} \cmidrule(lr){6-8} \cmidrule(lr){9-11}
&&BD-Rate&BD-Rate&BD-Rate&BD-Rate&BD-Rate&BD-Rate&BD-Rate&BD-Rate&BD-Rate\\
&&PSNR-Y&PSNR-U&PSNR-V&PSNR-Y&PSNR-U&PSNR-V&PSNR-Y&PSNR-U&PSNR-V\\
\midrule
\multirow{3}{*}{Netflix Sol Levante} 
&P1&0.04&3.22&2.95&1.10&4.67&4.45&2.18&6.19&5.84   \\
&P4&0.18&2.83&2.86&1.01&4.20&4.00&1.97&5.57&5.40   \\
&P7&0.12&2.78&2.77&1.09&4.07&4.02&1.98&5.64&5.34   \\
\midrule
\multirow{3}{*}{Netflix Meridian} 
&P1&-1.03&2.67&3.82&0.30&5.70&7.17&1.92&7.58&10.55 \\
&P4&-0.61&2.89&4.18&0.72&5.99&7.27&2.15&9.40&12.49 \\
&P7&0.01&2.95&3.80&1.05&5.90&6.73&2.51&8.91&11.53  \\
\midrule
\multirow{3}{*}{Netflix Nocturne} 
&P1&2.42&3.46&2.72&4.98&5.56&5.41&7.21&8.11&8.50   \\
&P4&3.82&4.18&3.59&6.34&6.86&6.70&8.89&9.37&10.05  \\
&P7&3.73&3.99&3.40&6.13&6.51&6.54&8.71&9.17&9.84   \\
\bottomrule
\end{tabular}}
\vspace{-6.5mm}
\end{table}

\clearpage

\subsection{Structural Similarity Index Measure (SSIM)}

\begin{table}[!h]
\vspace{-3mm}
\setstretch{0.65}
\caption{BD-Rate SSIM (\%) for each SFE configuration relative to No-SFE in H.265/HEVC 10-Bit 4:2:2 (ITE UHD Series A)}
\centering
\vspace{2mm}
\resizebox{14.5cm}{!}{\begin{tabular}{@{}lcccccccccc@{}}
\toprule
\multirow{3.5}{*}{Sequence Name} & \multirow{3.5}{*}{Preset} & \multicolumn{3}{c}{2 Ways} & \multicolumn{3}{c}{3 Ways} & \multicolumn{3}{c}{4 Ways}\\
\cmidrule(lr){3-5} \cmidrule(lr){6-8} \cmidrule(lr){9-11}
&&BD-Rate&BD-Rate&BD-Rate&BD-Rate&BD-Rate&BD-Rate&BD-Rate&BD-Rate&BD-Rate\\
&&SSIM-Y&SSIM-U&SSIM-V&SSIM-Y&SSIM-U&SSIM-V&SSIM-Y&SSIM-U&SSIM-V\\
\midrule
\multicolumn{11}{c}{\textbf{ITE UHD Series A 4K}}\\
\midrule
\multirow{3}{*}{a03\_TrainsC\_4K} 
&P1&1.94&2.01&0.98&3.32&2.48&0.74&5.68&3.11&-0.53         \\
&P4&2.46&2.13&1.02&3.76&2.67&1.90&4.89&2.99&1.56          \\
&P7&1.90&1.40&1.22&3.53&2.86&2.13&4.38&2.73&0.96          \\
\midrule
\multirow{3}{*}{a04\_Expressway\_4K} 
&P1&2.84&2.74&2.72&4.73&3.03&1.72&9.11&6.11&5.82          \\
&P4&3.60&3.49&2.27&7.48&6.67&3.80&10.41&8.99&6.34         \\
&P7&3.17&3.10&1.82&5.44&3.39&0.93&8.65&7.32&3.54          \\
\midrule
\multirow{3}{*}{a05\_SteelPlant\_4K} 
&P1&0.29&-0.41&0.10&0.54&-1.76&1.08&0.81&-0.05&2.27       \\
&P4&0.31&0.68&0.45&0.59&1.74&1.33&0.88&3.11&2.71          \\
&P7&0.28&0.57&0.46&0.59&1.73&1.41&0.88&2.85&2.71          \\
\midrule
\multirow{3}{*}{a06\_Festival\_4K} 
&P1&0.93&0.82&0.39&2.34&2.03&1.28&3.47&3.09&2.12          \\
&P4&1.21&1.01&0.45&2.48&2.03&1.51&3.66&3.16&2.28          \\
&P7&1.07&0.96&0.50&2.48&2.10&1.54&3.58&3.19&2.51          \\
\midrule
\multirow{3}{*}{a07\_River\_4K} 
&P1&0.49&4.91&0.31&0.89&1.60&-0.54&3.76&-3.69&-0.44       \\
&P4&0.62&3.23&0.48&0.99&-1.48&-0.22&3.99&-7.73&0.69       \\
&P7&0.30&2.60&-1.48&0.82&-3.46&-1.87&3.70&-1.05&-2.32     \\
\midrule
\multirow{3}{*}{a08\_JapaneseMaple\_4K} 
&P1&0.97&1.21&0.62&2.06&1.51&0.92&2.63&1.91&0.82          \\
&P4&0.94&1.11&0.92&1.24&0.99&0.44&1.52&1.57&0.88          \\
&P7&-0.04&0.31&0.00&1.97&1.65&1.24&3.10&2.53&1.87         \\
\midrule
\multirow{3}{*}{a09\_Maiko\_4K} 
&P1&0.00&0.55&-0.85&1.04&2.01&-0.30&2.68&3.27&1.53        \\
&P4&1.79&1.10&0.12&1.86&1.69&-0.46&3.23&2.34&0.77         \\
&P7&1.40&0.74&-0.14&3.18&2.46&1.85&4.89&5.25&3.35         \\
\midrule
\multirow{3}{*}{a10\_Umbrella\_4K} 
&P1&2.94&-1.60&1.82&4.27&0.64&3.24&1.14&-2.00&1.96        \\
&P4&2.67&2.92&0.78&0.31&1.86&-0.05&2.29&4.81&3.49         \\
&P7&-0.08&2.33&0.13&0.34&3.01&1.06&3.49&5.63&2.49         \\
\midrule
\multirow{3}{*}{a11\_LayeredKimono\_4K} 
&P1&-4.11&-0.60&-5.30&1.05&3.64&-1.05&-2.07&4.13&-3.27    \\
&P4&-1.82&1.12&-2.56&2.05&2.57&1.07&0.02&2.77&-0.42       \\
&P7&0.27&1.26&0.51&-1.64&1.94&-1.98&1.52&4.79&2.18        \\
\midrule
\multirow{3}{*}{a12\_Railcar\_4K} 
&P1&1.87&11.60&-1.41&2.39&-0.79&1.21&3.80&10.23&1.31      \\
&P4&1.94&21.67&2.98&2.73&21.14&2.86&4.18&26.75&7.15       \\
&P7&1.89&11.84&1.44&2.68&9.59&4.71&3.60&12.72&4.22        \\
\midrule
\multirow{3}{*}{\textbf{Average}} 
&P1&0.82&2.12&-0.06&2.26&1.44&0.83&3.10&2.61&1.16         \\
&P4&1.37&3.85&0.69&2.35&3.99&1.22&3.51&4.88&2.55          \\
&P7&1.02&2.51&0.45&1.94&2.53&1.10&3.78&4.60&2.15          \\
\midrule
\multicolumn{11}{c}{\textbf{ITE UHD Series A 8K}}\\
\midrule
\multirow{3}{*}{a01\_TrainsA\_8K} 
&P1&3.81&4.71&5.18&4.58&5.45&4.77&4.61&5.06&3.85          \\
&P4&3.32&5.33&4.19&4.13&5.96&4.38&6.40&8.90&8.22          \\
&P7&3.59&6.13&5.43&3.68&5.59&4.53&6.08&8.39&8.12          \\
\midrule
\multirow{3}{*}{a02\_TrainsB\_8K} 
&P1&-1.48&-0.88&-2.65&-0.74&-0.11&-1.76&-0.63&1.01&-0.09  \\
&P4&-0.47&-0.70&-1.50&-0.96&-1.15&-1.80&-0.80&-0.18&-0.76 \\
&P7&-0.47&-0.54&-1.43&-0.79&-1.06&-2.00&-0.53&0.27&-0.87  \\
\midrule
\multirow{3}{*}{a04\_Expressway\_8K} 
&P1&0.32&2.48&3.14&1.01&2.21&3.22&0.27&1.96&2.84          \\
&P4&-2.72&1.04&-1.03&-0.38&2.55&2.49&0.35&2.98&2.35       \\
&P7&-1.58&2.17&0.99&-0.26&1.88&1.76&0.17&2.21&0.95        \\
\midrule
\multirow{3}{*}{a05\_SteelPlant\_8K} 
&P1&0.25&0.33&-0.27&0.38&0.15&0.32&0.65&0.64&0.81         \\
&P4&0.31&0.27&-0.12&0.48&0.62&0.51&0.65&1.11&0.88         \\
&P7&0.25&0.19&-0.11&0.39&0.60&0.45&0.61&1.03&0.89         \\
\midrule
\multirow{3}{*}{a06\_Festival\_8K} 
&P1&0.22&0.44&-0.02&0.82&0.90&0.39&1.45&1.40&0.72         \\
&P4&0.26&0.59&0.07&0.90&1.10&0.49&1.59&1.68&0.98          \\
&P7&0.35&0.58&0.16&1.05&1.17&0.65&1.70&1.73&1.09          \\
\midrule
\multirow{3}{*}{a07\_River\_8K} 
&P1&0.14&-0.21&2.98&0.88&3.00&0.43&0.43&3.73&10.85        \\
&P4&0.28&-1.31&2.79&0.98&20.75&0.31&0.67&-1.34&10.45      \\
&P7&0.16&12.62&3.11&1.12&26.23&-0.24&0.53&24.64&11.18     \\
\midrule
\multirow{3}{*}{a08\_JapaneseMaple\_8K} 
&P1&0.80&2.01&1.59&0.72&1.81&1.44&0.97&2.35&2.18          \\
&P4&1.16&2.19&2.09&0.97&2.30&1.52&1.13&2.06&1.05          \\
&P7&0.01&0.24&2.78&0.71&0.44&2.36&0.55&0.00&1.19          \\
\midrule
\multirow{3}{*}{a09\_Maiko\_8K} 
&P1&3.71&1.58&-0.58&5.40&29.86&0.74&5.92&2.01&0.66        \\
&P4&3.23&-3.26&-1.87&3.51&2.01&-0.81&4.48&3.07&-0.26      \\
&P7&3.65&1.91&-0.41&3.91&3.39&-0.58&3.77&2.24&-0.18       \\
\midrule
\multirow{3}{*}{a10\_Umbrella\_8K} 
&P1&-2.92&1.32&-0.45&-3.26&2.04&-0.73&-4.34&1.02&-1.93    \\
&P4&-1.35&3.85&1.27&-0.98&3.73&1.04&0.50&7.50&3.39        \\
&P7&-0.35&2.27&0.15&-2.41&1.29&-0.87&-0.97&4.70&0.49      \\
\midrule
\multirow{3}{*}{a11\_LayeredKimono\_8K} 
&P1&-0.93&-2.41&-2.22&4.37&1.70&-4.63&7.97&5.09&-1.65     \\
&P4&-2.02&4.40&-2.23&-0.30&6.40&-0.33&0.64&5.62&-0.63     \\
&P7&-5.61&0.85&-5.96&-2.81&-0.55&-2.66&-5.05&1.31&-5.75   \\
\midrule
\multirow{3}{*}{a12\_Railcar\_8K} 
&P1&0.37&-3.25&-4.63&-0.26&-2.47&-4.34&-0.16&-3.71&-6.01  \\
&P4&0.74&-0.60&-3.91&0.43&-2.38&-5.24&0.48&-2.80&-6.51    \\
&P7&0.59&-3.73&-6.01&0.51&-3.55&-5.59&0.86&-3.09&-5.80    \\
\midrule
\multirow{3}{*}{\textbf{Average}} 
&P1&0.39&0.56&0.19&1.26&4.05&-0.01&1.56&1.87&1.11         \\
&P4&0.25&1.07&-0.02&0.80&3.81&0.23&1.46&2.60&1.74         \\
&P7&0.05&2.06&-0.12&0.46&3.22&-0.20&0.70&3.95&1.03        \\

\bottomrule
\end{tabular}}
\vspace{-6.5mm}
\end{table}

\begin{table}[!h]
\vspace{-3mm}
\setstretch{0.65}
\caption{BD-Rate SSIM (\%) for each SFE configuration relative to No-SFE in H.265/HEVC 10-Bit 4:2:2 (Netflix Chimera)}
\centering
\vspace{2mm}
\resizebox{14.5cm}{!}{\begin{tabular}{@{}lcccccccccc@{}}
\toprule
\multirow{3.5}{*}{Sequence Name} & \multirow{3.5}{*}{Preset} & \multicolumn{3}{c}{2 Ways} & \multicolumn{3}{c}{3 Ways} & \multicolumn{3}{c}{4 Ways}\\
\cmidrule(lr){3-5} \cmidrule(lr){6-8} \cmidrule(lr){9-11}
&&BD-Rate&BD-Rate&BD-Rate&BD-Rate&BD-Rate&BD-Rate&BD-Rate&BD-Rate&BD-Rate\\
&&SSIM-Y&SSIM-U&SSIM-V&SSIM-Y&SSIM-U&SSIM-V&SSIM-Y&SSIM-U&SSIM-V\\
\midrule
\multirow{3}{*}{1\_BAR\_SCENE} 
&P1&-0.08&1.02&-1.12&1.03&4.27&1.85&1.53&2.45&-0.20            \\
&P4&3.00&4.37&4.47&3.66&4.50&5.57&4.86&6.36&7.36               \\
&P7&2.08&-1.37&-1.48&2.22&0.70&1.38&3.85&1.75&1.23             \\
\midrule
\multirow{3}{*}{2\_DINNER\_SCENE} 
&P1&3.25&0.54&22.34&-7.71&-2.75&-0.08&-7.55&-1.30&-1.38        \\
&P4&6.30&17.70&-5.78&7.86&21.13&15.32&9.61&21.12&18.70         \\
&P7&6.48&8.66&15.96&8.51&16.06&26.21&8.67&9.95&13.22           \\
\midrule
\multirow{3}{*}{3\_DANCER} 
&P1&2.04&1.53&9.32&1.93&3.07&10.28&2.00&1.38&14.06             \\
&P4&-11.43&-25.10&46.85&-15.26&-22.19&56.96&-17.17&-9.16&75.49 \\
&P7&-12.70&-12.21&30.74&-14.02&-14.46&34.60&-13.37&-9.88&35.64 \\
\midrule
\multirow{3}{*}{4\_DANCERS\_COUPLE} 
&P1&4.18&-0.67&10.18&-2.33&-4.48&3.60&-1.06&-2.96&10.64        \\
&P4&-2.76&8.27&1.45&-0.25&13.60&7.63&-0.54&15.07&10.86         \\
&P7&3.35&17.43&-11.12&-0.33&16.21&6.79&5.49&18.44&11.65        \\
\midrule
\multirow{3}{*}{5\_DANCERS\_MONTAGE\_MIXED} 
&P1&2.38&-4.36&13.32&2.65&-7.46&6.39&2.81&-5.90&9.18           \\
&P4&3.46&1.12&4.57&5.80&1.63&12.05&4.22&2.71&14.50             \\
&P7&3.05&-1.95&5.52&3.95&-0.28&10.50&6.79&-0.76&10.94          \\
\midrule
\multirow{3}{*}{6\_ROLLERCOASTER\_SEQUENCE} 
&P1&3.98&2.49&1.34&7.22&5.94&4.20&9.73&8.54&6.48               \\
&P4&4.53&2.63&1.87&7.40&5.50&4.23&9.67&8.23&6.64               \\
&P7&3.88&2.29&1.74&7.19&5.49&4.52&9.60&8.32&6.64               \\
\midrule
\multirow{3}{*}{7\_ROLLERCOASTER\_POV} 
&P1&5.34&4.53&3.37&8.79&7.66&5.85&11.02&10.38&8.52             \\
&P4&5.54&4.57&3.87&9.17&8.20&6.99&11.03&10.41&8.80             \\
&P7&4.95&3.98&2.98&8.46&7.52&6.27&10.64&10.03&8.24             \\
\midrule
\multirow{3}{*}{8\_ROLLERCOASTER\_PASSENGER} 
&P1&3.29&1.81&1.04&3.74&1.57&0.79&6.68&4.35&3.59               \\
&P4&3.70&1.81&1.07&6.09&4.29&3.29&8.66&6.61&5.21               \\
&P7&3.79&2.33&1.68&5.87&4.27&3.27&8.42&6.62&5.38               \\
\midrule
\multirow{3}{*}{9\_TWIRL\_RIDE\_BOARDWALK} 
&P1&2.58&3.01&2.49&5.03&5.35&4.35&7.11&7.42&6.19               \\
&P4&3.04&3.33&2.71&5.52&5.62&4.72&7.81&7.94&6.99               \\
&P7&2.96&3.25&2.69&5.47&5.61&4.59&7.56&7.66&6.61               \\
\midrule
\multirow{3}{*}{10\_NETFLIX\_CARD\_TWIRL} 
&P1&3.09&2.54&2.27&4.90&4.19&3.70&6.98&5.90&5.52               \\
&P4&2.55&2.94&2.89&3.96&3.80&3.60&6.30&5.76&5.69               \\
&P7&3.02&3.13&3.18&4.92&5.14&5.03&6.95&6.50&6.57               \\
\midrule
\multirow{3}{*}{11\_SEASIDE\_AND\_PIER} 
&P1&3.61&2.80&2.64&5.21&3.82&3.04&6.65&5.93&4.65               \\
&P4&4.11&2.54&2.18&5.64&3.83&2.65&6.78&5.33&3.30               \\
&P7&4.02&2.36&2.69&5.58&3.01&2.60&6.92&5.36&3.93               \\
\midrule
\multirow{3}{*}{12\_WIND\_AND\_NATURE} 
&P1&1.25&-1.49&-1.93&1.42&0.85&1.11&2.08&10.73&1.35            \\
&P4&1.07&-2.17&-2.13&1.51&-0.92&-1.79&2.48&0.09&-1.09          \\
&P7&0.58&-3.18&-2.68&1.98&10.67&-0.04&2.67&-2.02&-3.00         \\
\midrule
\multirow{3}{*}{13\_MOUNTAIN\_VIEW\_W\_TILT} 
&P1&0.08&-13.17&1.63&3.11&-10.20&8.57&6.95&0.07&8.33           \\
&P4&-0.93&4.23&19.85&2.34&21.83&-15.24&6.70&23.32&-12.38       \\
&P7&3.04&12.12&20.54&5.24&21.77&-14.40&8.00&30.63&-11.90       \\
\midrule
\multirow{3}{*}{14\_MOUNTAIN\_VIEW\_PAN} 
&P1&2.02&1.05&0.05&3.29&1.67&1.36&6.40&5.60&6.32               \\
&P4&2.35&1.66&1.74&3.79&3.68&4.28&6.16&5.94&5.94               \\
&P7&1.52&-0.20&0.78&2.55&0.34&1.65&4.01&2.96&5.07              \\
\midrule
\multirow{3}{*}{15\_WALK\_LIKE\_A\_MAN} 
&P1&1.07&1.02&0.27&2.37&2.73&1.36&3.41&4.24&2.64               \\
&P4&2.60&0.00&0.14&3.15&1.90&1.73&2.70&1.09&0.64               \\
&P7&0.61&0.49&2.04&2.03&0.17&1.97&6.87&2.83&3.23               \\
\midrule
\multirow{3}{*}{16\_TODDLER\_AND\_FOUNTAIN} 
&P1&0.58&1.79&1.51&0.85&4.59&2.35&1.44&6.23&4.06               \\
&P4&0.65&2.45&1.72&1.08&2.85&2.43&1.68&6.81&4.34               \\
&P7&0.64&2.29&1.15&0.99&3.37&1.84&1.61&4.99&3.73               \\
\midrule
\multirow{3}{*}{17\_DRIVING\_POV} 
&P1&2.53&2.45&1.23&4.32&2.88&1.35&5.86&4.77&2.90               \\
&P4&2.36&2.34&1.34&4.19&3.54&2.13&6.18&5.89&3.97               \\
&P7&2.46&2.72&1.44&4.08&3.62&2.10&6.06&5.88&3.99               \\
\midrule
\multirow{3}{*}{18\_PLANET\_MOBILE} 
&P1&-0.11&-0.07&-3.52&0.67&0.09&0.42&1.25&1.71&4.11            \\
&P4&2.02&-0.42&-0.46&3.79&-7.24&3.67&3.34&-5.71&8.75           \\
&P7&2.06&-6.91&0.83&2.57&0.87&4.01&4.52&2.48&8.48              \\
\midrule
\multirow{3}{*}{19\_DOG\_PANTS} 
&P1&2.54&0.55&-1.31&4.96&2.37&0.85&6.32&5.15&2.96              \\
&P4&0.62&-0.46&-1.03&3.80&2.60&2.03&4.51&3.24&1.63             \\
&P7&2.24&0.87&-0.28&5.13&4.66&3.77&5.36&3.47&1.92              \\
\midrule
\multirow{3}{*}{20\_DOG\_BARKS} 
&P1&-2.48&-2.46&-2.96&-3.18&-5.79&-5.67&-1.09&-2.25&-2.48      \\
&P4&-1.11&0.00&-1.01&1.53&3.72&2.65&1.87&2.90&1.97             \\
&P7&-3.18&-2.08&-3.03&-2.08&-2.40&-3.42&0.59&0.83&0.00         \\
\midrule
\multirow{3}{*}{21\_RC\_AERIAL} 
&P1&2.05&1.84&-2.65&4.14&3.09&-1.56&5.82&4.45&-0.16            \\
&P4&0.18&-3.73&-3.25&2.46&-2.16&-1.37&3.95&-1.32&-0.35         \\
&P7&1.39&1.19&1.31&3.73&3.06&3.25&5.21&4.33&4.97               \\
\midrule
\multirow{3}{*}{22\_BASKEYBALL\_FREE\_THROW} 
&P1&6.55&5.50&6.48&8.59&9.33&9.08&10.19&11.38&11.44            \\
&P4&5.27&7.02&7.55&8.03&9.45&10.30&12.65&15.69&15.63           \\
&P7&3.90&5.09&7.08&7.52&8.79&9.99&11.37&14.57&15.78            \\
\midrule
\multirow{3}{*}{23\_BASKEYBALL\_GAME} 
&P1&5.12&4.17&3.38&8.51&7.62&6.03&11.05&10.85&8.39             \\
&P4&5.61&4.78&3.80&9.39&9.35&7.11&12.35&12.52&9.42             \\
&P7&5.97&5.59&4.41&9.35&9.20&7.08&12.35&12.53&9.43             \\
\midrule
\multirow{3}{*}{\textbf{Average}} 
&P1&2.39&0.71&3.02&3.02&1.76&3.01&4.59&4.31&5.09               \\
&P4&1.86&1.73&4.10&3.68&4.28&6.13&5.03&6.56&8.78               \\
&P7&2.00&2.00&3.83&3.52&4.93&5.37&5.66&6.41&6.60               \\
\bottomrule
\end{tabular}}
\vspace{-6.5mm}
\end{table}

\begin{table}[!h]
\vspace{-3mm}
\setstretch{0.65}
\caption{BD-Rate SSIM (\%) for each SFE configuration relative to No-SFE in H.265/HEVC 10-Bit 4:2:2 (Other)}
\centering
\vspace{2mm}
\resizebox{14.5cm}{!}{\begin{tabular}{@{}lcccccccccc@{}}
\toprule
\multirow{3.5}{*}{Sequence Name} & \multirow{3.5}{*}{Preset} & \multicolumn{3}{c}{2 Ways} & \multicolumn{3}{c}{3 Ways} & \multicolumn{3}{c}{4 Ways}\\
\cmidrule(lr){3-5} \cmidrule(lr){6-8} \cmidrule(lr){9-11}
&&BD-Rate&BD-Rate&BD-Rate&BD-Rate&BD-Rate&BD-Rate&BD-Rate&BD-Rate&BD-Rate\\
&&SSIM-Y&SSIM-U&SSIM-V&SSIM-Y&SSIM-U&SSIM-V&SSIM-Y&SSIM-U&SSIM-V\\
\midrule
\multirow{3}{*}{Netflix Sol Levante} 
&P1&3.56&3.39&2.99&4.79&4.65&3.86&5.88&5.81&4.88  \\
&P4&4.00&3.62&3.15&4.93&4.77&4.04&5.92&5.99&5.09  \\
&P7&3.96&3.43&3.05&4.96&4.72&3.99&6.19&6.00&5.07  \\
\midrule
\multirow{3}{*}{Netflix Meridian} 
&P1&1.30&-0.37&1.20&2.89&2.08&3.81&3.57&3.33&5.90 \\
&P4&2.80&0.18&1.54&4.12&2.65&4.09&5.12&4.88&7.36  \\
&P7&2.53&0.53&2.10&4.23&2.35&3.42&4.87&4.39&6.54  \\
\midrule
\multirow{3}{*}{Netflix Nocturne} 
&P1&2.42&3.05&2.49&4.85&4.46&4.67&7.19&6.20&7.54  \\
&P4&3.34&3.87&3.15&5.67&6.29&5.51&8.26&8.22&8.58  \\
&P7&3.31&3.91&3.18&5.56&5.85&5.65&8.02&8.16&8.64  \\
\bottomrule
\end{tabular}}
\vspace{-6.5mm}
\end{table}

\clearpage
\twocolumn
\subsection{Video Multimethod Assessment Fusion (VMAF) 4K -- No Enhancement Gain (NEG)}

\begin{table}[!h]
\vspace{-3mm}
\setstretch{0.65}
\caption{BD-Rate VMAF 4K NEG (\%) for each SFE configuration relative to No-SFE in H.265/HEVC 10-Bit 4:2:2 (ITE UHD Series A)}
\centering
\vspace{2mm}
\resizebox{6.5cm}{!}{\begin{tabular}{@{}lcccc@{}}
\toprule
Sequence Name & Preset & 2 Ways & 3 Ways & 4 Ways\\
\midrule
\multicolumn{5}{c}{\textbf{ITE UHD Series A 4K}}\\
\midrule
\multirow{3}{*}{a03\_TrainsC\_4K} 
&P1&2.28&5.24&8.35     \\
&P4&4.75&7.06&8.50     \\
&P7&3.70&5.67&7.22     \\
\midrule
\multirow{3}{*}{a04\_Expressway\_4K} 
&P1&2.72&4.54&7.85     \\
&P4&3.31&6.65&8.83     \\
&P7&2.91&5.40&8.18     \\
\midrule
\multirow{3}{*}{a05\_SteelPlant\_4K} 
&P1&0.18&0.47&0.71     \\
&P4&0.22&0.54&0.77     \\
&P7&0.21&0.48&0.75     \\
\midrule
\multirow{3}{*}{a06\_Festival\_4K} 
&P1&0.63&1.80&2.78     \\
&P4&0.84&1.93&2.90     \\
&P7&0.67&1.85&2.81     \\
\midrule
\multirow{3}{*}{a07\_River\_4K} 
&P1&0.49&0.77&1.48     \\
&P4&0.55&0.75&1.59     \\
&P7&0.43&0.67&1.39     \\
\midrule
\multirow{3}{*}{a08\_JapaneseMaple\_4K} 
&P1&0.40&1.77&2.16     \\
&P4&0.13&0.29&0.85     \\
&P7&-0.67&1.56&3.44    \\
\midrule
\multirow{3}{*}{a09\_Maiko\_4K} 
&P1&0.17&1.17&2.66     \\
&P4&2.32&2.22&3.47     \\
&P7&2.23&3.46&5.10     \\
\midrule
\multirow{3}{*}{a10\_Umbrella\_4K} 
&P1&0.06&1.83&1.05     \\
&P4&1.99&1.70&4.33     \\
&P7&0.79&2.17&3.61     \\
\midrule
\multirow{3}{*}{a11\_LayeredKimono\_4K} 
&P1&-4.46&0.36&-2.22   \\
&P4&-2.60&1.48&-0.69   \\
&P7&-1.23&-2.59&0.55   \\
\midrule
\multirow{3}{*}{a12\_Railcar\_4K} 
&P1&2.11&3.91&4.82     \\
&P4&2.65&4.93&6.64     \\
&P7&1.29&3.88&5.25     \\
\midrule
\multirow{3}{*}{\textbf{Average}} 
&P1&0.46&2.19&2.96     \\
&P4&1.42&2.76&3.72     \\
&P7&1.03&2.26&3.83     \\
\midrule
\multicolumn{5}{c}{\textbf{ITE UHD Series A 8K}}\\
\midrule
\multirow{3}{*}{a01\_TrainsA\_8K} 
&P1&4.45&5.26&5.20     \\
&P4&4.37&4.73&6.66     \\
&P7&4.32&4.27&6.22     \\
\midrule
\multirow{3}{*}{a02\_TrainsB\_8K} 
&P1&-0.60&-0.32&0.00   \\
&P4&0.34&0.12&0.16     \\
&P7&0.43&0.35&0.30     \\
\midrule
\multirow{3}{*}{a04\_Expressway\_8K} 
&P1&3.06&3.08&2.18     \\
&P4&0.75&2.38&3.20     \\
&P7&1.48&2.34&2.84     \\
\midrule
\multirow{3}{*}{a05\_SteelPlant\_8K} 
&P1&0.55&0.58&0.84     \\
&P4&0.55&0.74&0.82     \\
&P7&0.51&0.72&0.80     \\
\midrule
\multirow{3}{*}{a06\_Festival\_8K} 
&P1&0.41&1.04&1.57     \\
&P4&0.49&1.13&1.71     \\
&P7&0.58&1.25&1.78     \\
\midrule
\multirow{3}{*}{a07\_River\_8K} 
&P1&0.20&0.47&0.79     \\
&P4&0.20&0.54&0.75     \\
&P7&0.21&0.61&0.76     \\
\midrule
\multirow{3}{*}{a08\_JapaneseMaple\_8K} 
&P1&0.37&0.02&0.42     \\
&P4&1.11&0.56&0.82     \\
&P7&0.05&0.52&0.46     \\
\midrule
\multirow{3}{*}{a09\_Maiko\_8K} 
&P1&3.75&6.27&14.42    \\
&P4&4.66&4.60&3.06     \\
&P7&1.17&1.84&6.18     \\
\midrule
\multirow{3}{*}{a10\_Umbrella\_8K} 
&P1&-0.94&-1.29&-2.05  \\
&P4&-0.55&-0.41&0.86   \\
&P7&0.69&-0.21&0.47    \\
\midrule
\multirow{3}{*}{a11\_LayeredKimono\_8K} 
&P1&-6.02&2.81&6.96    \\
&P4&-4.20&-2.87&-0.95  \\
&P7&-8.08&-8.74&-7.49  \\
\midrule
\multirow{3}{*}{a12\_Railcar\_8K} 
&P1&1.23&1.20&1.04     \\
&P4&1.34&1.49&1.37     \\
&P7&1.56&1.80&1.57     \\
\midrule
\multirow{3}{*}{\textbf{Average}} 
&P1&0.59&1.74&2.85     \\
&P4&0.82&1.18&1.68     \\
&P7&0.27&0.43&1.26     \\
\bottomrule
\end{tabular}}
\vspace{-6.5mm}
\end{table}

\begin{table}[!h]
\vspace{-3mm}
\setstretch{0.65}
\caption{BD-Rate VMAF 4K NEG (\%) for each SFE configuration relative to No-SFE in H.265/HEVC 10-Bit 4:2:2 (Netflix Chimera)}
\centering
\vspace{2mm}
\resizebox{6.6cm}{!}{\begin{tabular}{@{}lcccc@{}}
\toprule
Sequence Name & Preset & 2 Ways & 3 Ways & 4 Ways\\
\midrule
\multirow{3}{*}{1\_BAR\_SCENE} 
&P1&-0.85&-0.07&0.24  \\
&P4&3.02&2.77&4.18    \\
&P7&1.44&0.81&2.88    \\
\midrule
\multirow{3}{*}{2\_DINNER\_SCENE} 
&P1&6.25&-18.65&-19.18\\
&P4&10.26&12.21&11.44 \\
&P7&9.67&9.46&12.10   \\
\midrule
\multirow{3}{*}{3\_DANCER} 
&P1&17.07&17.76&19.31 \\
&P4&13.09&10.47&10.09 \\
&P7&8.14&8.85&10.93   \\
\midrule
\multirow{3}{*}{4\_DANCERS\_COUPLE} 
&P1&12.01&-2.40&-0.03 \\
&P4&6.32&8.04&9.40    \\
&P7&6.63&8.81&9.31    \\
\midrule
\multirow{3}{*}{5\_DANCERS\_MONTAGE\_MIXED} 
&P1&26.66&29.52&33.32 \\
&P4&12.94&15.73&17.07 \\
&P7&11.49&14.84&16.19 \\
\midrule
\multirow{3}{*}{6\_ROLLERCOASTER\_SEQUENCE} 
&P1&4.50&7.41&9.83    \\
&P4&4.94&7.58&9.77    \\
&P7&4.19&7.29&9.75    \\
\midrule
\multirow{3}{*}{7\_ROLLERCOASTER\_POV} 
&P1&6.08&9.45&11.71   \\
&P4&5.86&9.41&11.53   \\
&P7&5.44&8.88&11.26   \\
\midrule
\multirow{3}{*}{8\_ROLLERCOASTER\_PASSENGER} 
&P1&3.68&3.96&6.12    \\
&P4&3.90&6.05&8.16    \\
&P7&4.00&5.82&8.03    \\
\midrule
\multirow{3}{*}{9\_TWIRL\_RIDE\_BOARDWALK} 
&P1&3.04&5.26&7.04    \\
&P4&3.13&5.35&7.23    \\
&P7&3.07&5.30&7.05    \\
\midrule
\multirow{3}{*}{10\_NETFLIX\_CARD\_TWIRL} 
&P1&3.65&5.28&7.02    \\
&P4&2.68&3.76&5.92    \\
&P7&3.21&4.79&6.63    \\
\midrule
\multirow{3}{*}{11\_SEASIDE\_AND\_PIER} 
&P1&3.09&4.61&5.78    \\
&P4&3.64&4.94&5.80    \\
&P7&3.49&4.85&5.74    \\
\midrule
\multirow{3}{*}{12\_WIND\_AND\_NATURE} 
&P1&0.75&6.19&65.69   \\
&P4&0.64&0.81&1.68    \\
&P7&0.01&0.41&1.12    \\
\midrule
\multirow{3}{*}{13\_MOUNTAIN\_VIEW\_W\_TILT} 
&P1&0.51&2.32&5.39    \\
&P4&-1.38&1.50&4.63   \\
&P7&3.22&4.55&6.14    \\
\midrule
\multirow{3}{*}{14\_MOUNTAIN\_VIEW\_PAN} 
&P1&0.56&2.78&4.23    \\
&P4&0.23&1.12&3.33    \\
&P7&2.25&2.90&4.03    \\
\midrule
\multirow{3}{*}{15\_WALK\_LIKE\_A\_MAN} 
&P1&2.79&4.82&5.04    \\
&P4&2.32&3.54&3.94    \\
&P7&5.42&3.78&7.80    \\
\midrule
\multirow{3}{*}{16\_TODDLER\_AND\_FOUNTAIN} 
&P1&0.79&1.33&2.13    \\
&P4&0.97&1.90&2.36    \\
&P7&0.99&1.55&2.56    \\
\midrule
\multirow{3}{*}{17\_DRIVING\_POV} 
&P1&3.06&5.17&6.70    \\
&P4&2.86&4.99&6.74    \\
&P7&3.01&4.86&6.62    \\
\midrule
\multirow{3}{*}{18\_PLANET\_MOBILE} 
&P1&-0.44&2.48&13.05  \\
&P4&-4.77&-1.97&0.87  \\
&P7&22.50&27.19&30.67 \\
\midrule
\multirow{3}{*}{19\_DOG\_PANTS} 
&P1&3.20&5.69&6.90    \\
&P4&2.71&5.05&5.89    \\
&P7&4.88&7.27&7.74    \\
\midrule
\multirow{3}{*}{20\_DOG\_BARKS} 
&P1&2.32&3.21&2.63    \\
&P4&3.45&6.38&7.59    \\
&P7&0.89&2.64&5.56    \\
\midrule
\multirow{3}{*}{21\_RC\_AERIAL} 
&P1&2.64&4.79&6.60    \\
&P4&2.00&4.06&5.81    \\
&P7&2.13&4.39&5.89    \\
\midrule
\multirow{3}{*}{22\_BASKEYBALL\_FREE\_THROW} 
&P1&5.44&9.02&9.48    \\
&P4&5.81&7.89&12.04   \\
&P7&4.95&8.47&12.38   \\
\midrule
\multirow{3}{*}{23\_BASKEYBALL\_GAME} 
&P1&5.70&9.25&11.55   \\
&P4&6.44&10.29&13.04  \\
&P7&6.99&10.31&12.97  \\
\midrule
\multirow{3}{*}{\textbf{Average}} 
&P1&4.89&5.18&9.59    \\
&P4&3.96&5.73&7.33    \\
&P7&5.13&6.87&8.84    \\

\bottomrule
\end{tabular}}
\vspace{-6.5mm}
\end{table}

\begin{table}[!h]
\vspace{1mm}
\setstretch{0.65}
\caption{BD-Rate VMAF 4K NEG (\%) for each SFE configuration relative to No-SFE in H.265/HEVC 10-Bit 4:2:2 (Other)}
\centering
\vspace{1mm}
\resizebox{6.5cm}{!}{\begin{tabular}{@{}lcccc@{}}
\toprule
Sequence Name & Preset & 2 Ways & 3 Ways & 4 Ways\\
\midrule
\multirow{3}{*}{Netflix Sol Levante} 
&P1&3.45&4.74&5.82  \\
&P4&3.85&4.67&5.71  \\
&P7&3.84&4.92&5.90  \\
\midrule
\multirow{3}{*}{Netflix Meridian} 
&P1&-0.08&0.90&2.02 \\
&P4&0.94&1.54&2.93  \\
&P7&1.10&1.78&2.95  \\
\midrule
\multirow{3}{*}{Netflix Nocturne} 
&P1&3.02&5.21&7.50  \\
&P4&3.62&5.79&8.22  \\
&P7&3.74&5.80&8.17  \\
\bottomrule
\end{tabular}}
\vspace{-6.5mm}
\end{table}

\end{document}